\documentclass[twocolumn,english,prl,aps,superscriptaddress,floatfix,showpacs]{revtex4-1}
\usepackage{textcomp}
\usepackage[latin9]{inputenc}
\usepackage{xcolor}
\usepackage{float}
\usepackage{amsmath}
\usepackage{amssymb}
\usepackage{graphicx}

\makeatletter

\providecommand{\tabularnewline}{\\}

\usepackage{dcolumn}
\usepackage{bm}
\usepackage{upgreek}
\usepackage[normalem]{ulem}
\usepackage{dsfont}
\usepackage[english]{babel}
\usepackage{babel}

\makeatother

\usepackage{babel}
\begin{document}
\title{{Interaction induced moir\'e systems in twisted bilayer optical lattices}}
\author{Jian-Hua Zeng}
\affiliation{Key Laboratory of Atomic and Subatomic Structure and Quantum Control
(Ministry of Education), Guangdong Basic Research Center of Excellence
for Structure and Fundamental Interactions of Matter, School of Physics,
South China Normal University, Guangzhou 510006, China}
\affiliation{School of Physics, Sun Yat-sen University, Guangzhou 510275, China}
\author{Qizhong Zhu}
\email{qzzhu@m.scnu.edu.cn}
\affiliation{Key Laboratory of Atomic and Subatomic Structure and Quantum Control
(Ministry of Education), Guangdong Basic Research Center of Excellence
for Structure and Fundamental Interactions of Matter, School of Physics,
South China Normal University, Guangzhou 510006, China}
\affiliation{Guangdong Provincial Key Laboratory of Quantum Engineering and Quantum
Materials, Guangdong-Hong Kong Joint Laboratory of Quantum Matter,
South China Normal University, Guangzhou 510006, China}
\author{Liang He}
\email{liang.he@scnu.edu.cn}

\affiliation{Key Laboratory of Atomic and Subatomic Structure and Quantum Control
(Ministry of Education), Guangdong Basic Research Center of Excellence
for Structure and Fundamental Interactions of Matter, School of Physics,
South China Normal University, Guangzhou 510006, China}
\affiliation{Guangdong Provincial Key Laboratory of Quantum Engineering and Quantum
Materials, Guangdong-Hong Kong Joint Laboratory of Quantum Matter,
South China Normal University, Guangzhou 510006, China}

\begin{abstract}
Moir\'e related physics in twisted bilayer two-dimensional (2D) materials
has attracted widespread interest in condensed matter physics. Simulation
of moir\'e related physics in cold atom platform is expected to outperform
the 2D materials thanks to its advantage of higher tunablility. Here,
we demonstrate that, the cold atom platform enables a new mechanism
of moir\'e lattice formation, { induced by interlayer interaction with
intrinsic ``dynamical'' character,} in contrast to conventional
moir\'e lattice induced by ``static'' ways such as single-particle
interlayer tunneling. Specifically, we consider a twisted bilayer
Bose-Hubbard model with vanishing interlayer tunneling, and the bilayer
is solely coupled through interlayer interaction that originates from
contact interaction of atoms. We find that this system hosts a plethora
of novel phases unique to this dynamical lattice, including a variety
of Mott insulator (MI) and superfluid (SF) phases either preserving
or breaking moir\'e lattice symmetry, phases with one layer in SF and
the other in MI, ``interlocked'' MI, and self-localized phases at
\textit{commensurate} twist angles, which exhibits the characteristics
of Bose glass and quasi-many-body localization in the absence of (quasi)disorder
or quasicrystalline lattices. Our prediction can be readily observed
in current experimental setup of twisted bilayer optical lattices,
opening up new avenues for exploring the rich physics of interaction
induced moir\'e systems in cold atoms. 
\end{abstract}
\maketitle

\section{Introduction}

Moir\'e physics has been extensively studied in the field of condensed
matter physics in recent years. One prominent feature that stimulates
tremendous interest is the emergence of flat band by simply tuning
the twist angle of bilayer two-dimensional (2D) materials, which offers
an intriguing platform for exploring phases that electron-electron
interaction plays a significant role, such as correlated insulators
\cite{cao_Correlated_2018,li_Continuous_2021}, superconductivity
\cite{cao_Unconventional_2018}, magnetism \cite{gong_Discovery_2017,huang_Layerdependent_2017,chen_Tunable_2020}
and generalized Wigner crystals \cite{regan_Mott_2020a,li_Imaging_2021a}.
On the other hand, topological phases also emerge in moir\'e superlattices,
e.g., quantum anomalous Hall insulators (QAHI) \cite{serlin_Intrinsic_2020,li_Quantum_2021,zhou_Bilayer_2021a,wu_PRX},
$\mathbb{Z}_{2}$ topological insulators \cite{wu_Topological_2019,zhao_Realization_2024},
high-order topological insulators \cite{park_HigherOrder_2019}, and
fractional QAHI \cite{cai_Signatures_2023,park_Observation_2023a,zeng_Thermodynamic_2023,xu_Observation_2023,lu2024fractional}.
The rich many-body and topological physics observed in bilayer 2D
materials have clearly demonstrated that twisting serves as a novel
and effective means for manipulating material properties and unlocking
diverse and fascinating quantum phenomena \cite{kennes_Moire_2021}.

Exploring the moir\'e physics in other platforms is appealing to fully
unleash the power of twisting. One prospective platform is cold atoms,
where optical lattice has been a faithful simulation of lattice physics
in condensed matter \cite{Bloch2008ultracold}. A variety of theoretical
proposals have been put forward \cite{gonzalez-tudela_Cold_2019,salamon_Simulating_2020,luo_SpinTwisted_2021,paul2023particle,wan2024fractal,wang2024threedimensional,madronero2023dynamic,madronero2024localized},
where twisted moir\'e physics can be either simulated using bilayer
optical lattice or using pseudospin-1/2 atomic species. Very recently,
exciting experimental progress has been made \cite{meng_Atomic_2023},
which has turned long-standing speculation into reality. Compared
to 2D materials, cold atom platforms offer significant advantages
\cite{Bloch2008ultracold,Cheng2010Feshbach,Gross_Nat_Phys_2021},
as they allow for the engineering of diverse monolayer lattice structures,
dynamic tuning of twist angles, and straightforward adjustment of
interlayer coupling. Moreover, being free from lattice relaxation,
strain, and disorder, these platforms support pristine twisted bilayer
models, allowing direct comparisons between theories and experiments.

So far, the moir\'e lattices observed in both 2D materials and cold
atom studies share a common characteristic: they are ``static'',
in the sense that their form is uniquely determined by interlayer
tunneling, structural relaxations, or external controls such as dielectric
screening \cite{bistritzer2011moire,wang2017interlayer,ruiz2019interlayer,li_Imaging_2021,xu2021creation,kim2024electrostatic,zhang2024engineering,gu2024remote,he2024dynamically}.
Here in this work, we propose a distinct mechanism of realizing moir\'e
lattice purely induced by particle-particle interaction from two layers
or two pseudospins, which we refer to as ``dynamical'' moir\'e lattice.
{ With vanishing interlayer coupling at single-particle level, the
two layers are completely isolated, and thus there is no moir\'e lattice
at this level.} For each layer/pseudospin of lattice, the effect
of the other layer is brought by particle-particle interaction, which
relies on particle density, a dynamical variable \footnote{Therefore, each layer can feel dynamical feedback from the other layer
and the ultimate density distribution is a compromise between the
two layers. This is to be contrasted with dynamical control of moir\'e
lattice already realized in 2D materials, where the moir\'e potential
in target layer is uniquely imprinted by substrate layers and has
no feedback on the substrate layers \cite{kim2024electrostatic,zhang2024engineering,gu2024remote,he2024dynamically}}. 
Realizing such a dynamical moir\'e system in twisted 2D materials is
challenging, but it can be easily achieved in modern cold atom platforms
by simply turning off interspin coupling in the twisted spin-dependent
lattices \cite{meng_Atomic_2023}.

\begin{figure*}[t]
\begin{centering}
\includegraphics[width=6.85in]{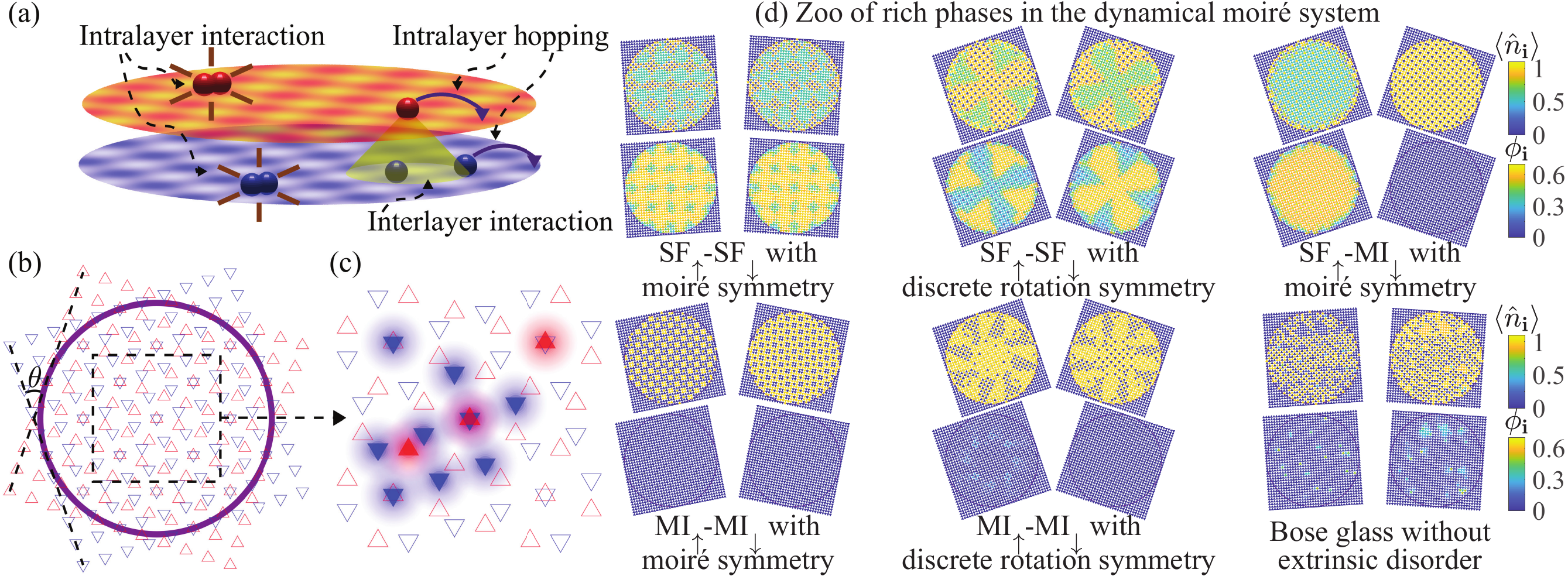} 
\par\end{centering}
\caption{(a) Schematic illustration of the dynamical moir\'e system. (b) Sketch
of a twisted bilayer optical lattice at the twisted angle $\theta=2\arctan(1/3)=36.87^{\circ}$.
Red (blue) triangles correspond to lattice sites of $\uparrow$ ($\downarrow$)
layer. The purple circle denotes the external cylinder box trap. (c)
Schematic illustration of the origin of the interlayer interaction.
Filled triangles denote the lattice sites occupied by atoms. The overlap
between atomic wave functions (denoted by the translucent circles)
of atoms on different layers gives rise to the interlayer interaction.
(d) Zoo of rich phases in the dynamical moir\'e system. The left (right)
column of each subplot corresponds to the $\uparrow(\downarrow)$-layer,
respectively.{} From left to right, up to down, ($\theta$, $\rho_{\uparrow}$,
$\rho_{\downarrow}$, $g_{\uparrow\downarrow}/g_{\sigma\sigma}$,
$J/U(g_{\sigma\sigma})$) assume the values ($7.63^{\circ}$, $3/5$,
$3/5$, $0.5$, $0.06$), ($36.87^{\circ}$, $4/5$, $4/5$, $0.9$,
$0.048$), ($36.87^{\circ}$, $7/10$, $4/5$, $0.9$, $0.04$), ($22.62^{\circ}$,
$2/5$, $3/5$, $0.9$, $0.024$), ($36.87^{\circ}$, $4/5$, $4/5$,
$0.3$, $0.008$), and ($7.63^{\circ}$, $2/5$, $3/5$, $0.5$, $0.024$),
respectively. }
\label{fig:model_and_overview} 
\end{figure*}

Specifically, we consider two species of ultracold atoms loaded in
a twisted bilayer optical lattice and find that, an emergent moir\'e
lattice from particle interaction can form, hosting much more rich
phases than static lattice counterpart (see Fig.~\ref{fig:model_and_overview}),
including phases with and without moir\'e lattice symmetry and Mott
insulator (MI) phases induced by interlayer interaction even for partial
filling. More interestingly, for total unit filling, the ground state
can become an ``interlocked'' MI with spontaneously formed moir\'e
lattice. For a general total filling, the {Bose glass (BG) }phase
exists, suggesting that the ground state tends to localize due to
the interplay between twisting and interlayer interaction, even at
commensurate twist angles.

\section{System and model}

Motivated by the recent realization of the twisted bilayer optical
lattices \cite{meng_Atomic_2023}, we consider a closely related but
\emph{simpler} experimental setup, namely, a twisted bilayer optical
lattice system \textit{without} the additional microwave field for
realizing the interlayer coupling in the recent experiments \cite{meng_Atomic_2023}.
More specifically, the system consists of two species of bosonic atoms
that correspond to two hyperfine states (for instance, $^{87}\mathrm{Rb}$
atoms in hyperfine states $|F=1,m_{F}=1\rangle$ and $|F=2,m_{F}=0\rangle$
considered in Ref.~\cite{meng_Atomic_2023}), with each species loaded
in one layer of a spin-dependent twisted bilayer square optical lattice,
respectively {[}see Fig.~\ref{fig:model_and_overview}(a){]}.

The physics of this system at low filling can be described by a Bose-Hubbard
type model within the lowest band approximation {[}see Appendix~\ref{sec:System_model}
for a detailed derivation{]}, 
\begin{align}
\hat{H}= & \sum_{\sigma=\uparrow,\downarrow}\left[-J\sum_{\langle\mathbf{i}_{\sigma},\mathbf{j}_{\sigma}\rangle}\hat{b}_{\mathbf{i_{\sigma}}}^{\dagger}\hat{b}_{\mathbf{j_{\sigma}}}+\sum_{\mathbf{i}_{\sigma}}\frac{U(g_{\sigma\sigma})}{2}\hat{n}_{\mathbf{i}_{\sigma}}(\hat{n}_{\mathbf{i}_{\sigma}}-1)\right]\nonumber \\
 & +\sum_{\sigma=\uparrow,\downarrow}\sum_{\langle\mathbf{i}_{\sigma},\mathbf{j}_{\sigma}\rangle}\frac{U_{\mathbf{i}_{\sigma}\mathbf{j}_{\sigma}}(g_{\sigma\sigma})}{2}\hat{n}_{\mathbf{i}_{\sigma}}\hat{n}_{\mathbf{j}_{\sigma}}\nonumber \\
 & +\sum_{\langle\mathbf{i}_{\uparrow},\mathbf{j}_{\downarrow}\rangle}U_{\mathbf{i}_{\uparrow}\mathbf{j}_{\downarrow}}(g_{\uparrow\downarrow},\theta)\hat{n}_{\mathbf{i}_{\uparrow}}\hat{n}_{\mathbf{j}_{\downarrow}},\label{eq:Hamiltonian}
\end{align}
where $\sigma=\uparrow,\downarrow$ is the index of the atom species/lattice
layer, $\hat{b}_{\mathbf{i_{\sigma}}}^{\dagger}(\hat{b}_{\mathbf{i}_{\sigma}})$
is the creation (annihilation) operator at site $\mathbf{i}$ of the
lattice layer $\sigma$ in the Wannier basis, $\hat{n}_{\mathbf{i}_{\sigma}}\equiv\hat{b}_{\mathbf{i}_{\sigma}}^{\dagger}\hat{b}_{\mathbf{i}_{\sigma}}$
is the corresponding particle number operator, and $\langle\cdots\rangle$
denotes nearest-neighbor lattice sites.

The first two parts of $\hat{H}$ basically assume the form of the
conventional Bose-Hubbard model, which consists of the hopping term
of different species of atoms in its own layer with the hopping amplitude
being $J$ and the interaction terms originating from the intra-species
contact interactions with strength $g_{\sigma\sigma}$ between atoms
on the same layer. For the latter, we not only take into account the
conventional on-site interaction term with the strength denoted by
$U(g_{\sigma\sigma})$, but also the nearest-neighbor interaction
terms with the strengths denoted by $U_{\mathbf{i}_{\sigma}\mathbf{j}_{\sigma}}(g_{\sigma\sigma})$,
and treat the inter-species contact interaction on an equal footing.

The third part of $\hat{H}$ captures the distinct way that the twist
imposes its physical influences even without the interlayer tunneling.
It originates from the finite overlaps between wave functions of atoms
on different layers with their inter-species contact interaction being
$g_{\uparrow\downarrow}$ {[}see Fig.~\ref{fig:model_and_overview}(c){]},
and consists of interlayer interaction terms $U_{\mathbf{i}_{\uparrow}\mathbf{j}_{\downarrow}}(g_{\uparrow\downarrow},\theta)\hat{n}_{\mathbf{i}_{\uparrow}}\hat{n}_{\mathbf{j}_{\downarrow}}$,
with the interaction strength $U_{\mathbf{i}_{\uparrow}\mathbf{j}_{\downarrow}}(g_{\uparrow\downarrow},\theta)$
assuming the twist angle $\theta$ dependence. {As the overlaps of
the wave functions depend directly on the spatial distance of sites
$\mathbf{i}_{\uparrow}$ and $\mathbf{j}_{\downarrow}$, the interaction
strength $U_{\mathbf{i}_{\uparrow}\mathbf{j}_{\downarrow}}(g_{\uparrow\downarrow},\theta)$
has explicit dependence on both spatial distance and twist angle $\theta$.}
{While many studies have investigated multicomponent Bose-Hubbard
models in the context of ultracold atoms, our study focuses on the
largely unexplored many-body physics of two-component ultracold bosons
in twisted bilayer optical lattices. The rich physics arising from
the twist-dependent interlayer interaction $U_{\mathbf{i}_{\uparrow}\mathbf{j}_{\downarrow}}(g_{\uparrow\downarrow},\theta)$
in such systems has not yet been systematically studied.}

We investigate this system by calculating its ground state via the
bosonic Gutzwiller variational approach \cite{Krauth_PRB_1992,Jaksch_PRL_1998,Lanata_PRB_2012}
(see Appendix~\ref{sec:System_model} for technical details). We
have chosen the twist angle $\theta=2\arctan(\bar{m}/\bar{n})$ with
$\bar{m},\bar{n}$ being two natural numbers \cite{wang_Localization_2020,meng_Atomic_2023},
i.e., the commensurate condition for twisted square lattice, and focus
on the case with $g_{\uparrow\uparrow}=g_{\downarrow\downarrow}$.
We perform the numerical simulation on a $37\times37$ twisted bilayer
optical lattice with an external 2D cylinder box trap \cite{Navon_Nat_Phys_2021}
whose radius is 18 lattice constants (denoted by the solid circles
in the figures). {We additionally investigate the finite-size effects
and find no significant differences between the results here and those
on a larger system (see the Appendix~\ref{sec:Finite_size}).} In
the following, we focus on the most interesting case of fractional
filling for both layers, and reserve the results of integer filling
in the Appendix~\ref{sec:integer_filling} (the impact of interlayer
interaction is not as dramatic in the latter case).

Before presenting the numerical results, we first notice that in conventional
moir\'e systems \cite{bistritzer2011moire,wang2017interlayer,ruiz2019interlayer,meng_Atomic_2023,wu_PRX},
{ the moir\'e lattices in twisted bilayers are induced by single-particle
interlayer tunneling. Particle interactions can be further considered
within the framework of these preformed moir\'e lattices. In sharp contrast,
here, the twisted bilayers are completely isolated at single-particle
level, and the moir\'e lattice arises purely from the interlayer interaction
terms of the Hamiltonian (\ref{eq:Hamiltonian}).} This novel mechanism
of moir\'e formation could make the system feature not only similar
phases in the conventional moir\'e system, but also many intriguing
phases without counterpart in the latter {[}see Fig.~\ref{fig:model_and_overview}(d)
for an overview{]}.

\begin{figure*}[t]
\begin{centering}
\includegraphics[width=6.85in]{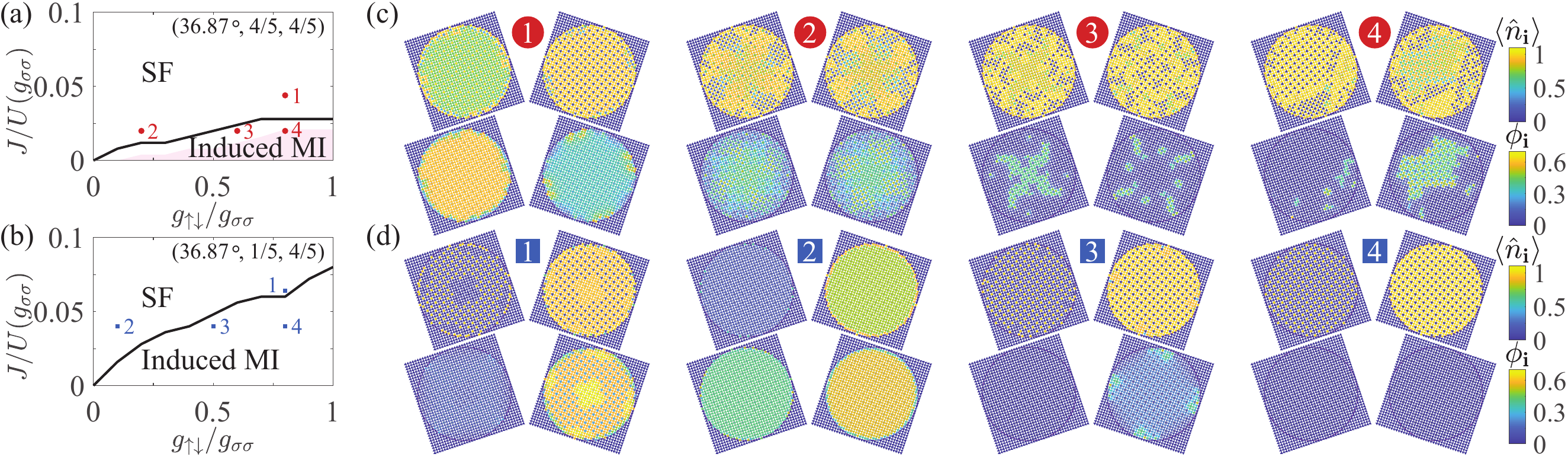} 
\par\end{centering}
\caption{ Phase diagrams and typical configurations at the balanced and imbalanced
fillings. (a) Phase diagram with respect to the interlayer interaction
strength $g_{\uparrow\downarrow}/g_{\sigma\sigma}$ and the hopping
amplitude $J/U(g_{\sigma\sigma})$ at the balanced filling with $\rho_{\uparrow}=4/5$
and $\rho_{\downarrow}=4/5$ (twist angle is fixed at $\theta=36.87^{\circ}$).
{The magenta region of the phase diagram contains the BG phase.}
(b) Similar to (a), but at the imbalanced filling with $\rho_{\uparrow}=1/5$
and $\rho_{\downarrow}=4/5$. (c) and (d) Real-space particle density
and superfluid order parameter distributions of the system that correspond
to the red dots and the blue squares in (a) and (b), respectively. }
\label{fig:balanced_and_imbalanced_filling} 
\end{figure*}

\section{Results}

\subsection{Conventional moir\'e physics without interlayer tunneling}

In conventional bosonic moir\'e systems, the interlayer tunneling causes
layer hybridization of single particle orbitals, and thus particle
filling on the hybridized orbitals results in phases with two layer
sharing the same character. Indeed, as observed in recent experiments
\cite{meng_Atomic_2023}, it can accommodate SF$_{\uparrow}$-SF$_{\downarrow}$
and MI$_{\uparrow}$-MI$_{\downarrow}$ (denoting SF/MI phase in $\uparrow/\downarrow$
layer), with the particle density assuming the moir\'e lattice symmetry.
In fact, these observations can be easily explained by treating the
system as if the atoms on one layer experience an additional static
potential from the other layer of the lattice \cite{meng_Atomic_2023}.
Speculating along this line, we notice that although there is no interlayer
tunneling in the dynamical moir\'e system studied here, atoms residing
on one layer can nevertheless feel an additional potential provided
by the atoms residing on the other layer via the interlayer interactions
$\sum_{\langle\mathbf{i}_{\uparrow},\mathbf{j}_{\downarrow}\rangle}U_{\mathbf{i}_{\uparrow}\mathbf{j}_{\downarrow}}(g_{\uparrow\downarrow},\theta)\hat{n}_{\mathbf{i}_{\uparrow}}\hat{n}_{\mathbf{j}_{\downarrow}}$.
This indicates the system may accommodate similar conventional moir\'e
physics even \emph{without} the interlayer tunneling.

In Fig.~\ref{fig:balanced_and_imbalanced_filling}(a), we map out
the phase diagram of the system in a balanced filling case with $\rho_{\uparrow}=\rho_{\downarrow}=4/5$
($\rho_{\sigma}\equiv\langle\sum_{\mathbf{i}_{\sigma}}\hat{n}_{\mathbf{i}_{\sigma}}\rangle/N_{\sigma}^{\mathrm{lat}}$
with $N_{\sigma}^{\mathrm{lat}}$ being the number of lattice sites
of the $\sigma$-layer within the cylinder trap). In the relatively
large hopping regime, we can see that the system indeed support SF$_{\uparrow}$-SF$_{\downarrow}$
phase (denoted as SF in the phase diagrams) similar to conventional
static moir\'e systems, with both layers manifesting the SF order parameter
$\phi_{\mathbf{i}_{\sigma}}\equiv\langle\hat{b}_{\mathbf{i}_{\sigma}}\rangle$
and density distribution $\langle\hat{n}_{\mathbf{i}_{\sigma}}\rangle$
assuming the moir\'e lattice symmetry {{[}see Fig.~\ref{fig:balanced_and_imbalanced_filling}(c1){]}}.
While in the relatively small hopping regime, we notice that despite
the filling factor on each layer being non-integer, increasing the
inter-species interaction strength $g_{\uparrow\downarrow}$ can drive
the system into the induced MI phase characterized by regions with
integer-filled particles and vanishing SF order parameter in either
layer{{} {[}see Fig.~\ref{fig:balanced_and_imbalanced_filling}(c4){]}},
including MI$_{\uparrow}$-SF$_{\downarrow}$, MI$_{\uparrow}$-MI$_{\downarrow}$
and SF$_{\uparrow}$-MI$_{\downarrow}$ phases. This is in sharp contrast
to the single-component Bose gases in optical lattices, where the
MI can only exist at integer fillings.

The induced MI phase occurs because increased $g_{\uparrow\downarrow}$
enhances the effective potential experienced by atoms on one layer
imposed by atoms on the other layer, causing localization. This scenario
is reminiscent of conventional bosonic moir\'e systems, where increasing
the interlayer tunneling makes one layer impose a stronger additional
static lattice potential on the other layer, leading to stronger localization
\cite{meng_Atomic_2023}. However, it's important to recognize a key
difference here: the effective potential is a dynamical one determined
by the dynamical variables, and hence does not necessarily share the
same spatial symmetry of the underlying optical lattice. For balanced
filling case, indeed, we have not found MI phase with both layer {assuming}
the moir\'e lattice symmetry. Instead, as shown in the {{} Figs.~\ref{fig:balanced_and_imbalanced_filling}(c2)
and (c3)}, we find this dynamical moir\'e system can manifest rich
discrete rotational symmetric MI and SF phases with no counterparts
in the conventional bosonic moir\'e systems. However, away from the
balanced filling, we indeed find the system can support MI phase with
density distributions on both layers assuming the moir\'e lattice symmetry.
For instance, {as shown in the Fig.~\ref{fig:balanced_and_imbalanced_filling}(d4)},
this kind of moir\'e MI phase can emerge in the induced MI regime of
the system at an imbalanced filling $\rho_{\uparrow}=1/5,\rho_{\downarrow}=4/5$.
{{} In this case, $\uparrow$-particles occupy overlapping sites
while $\downarrow$-particles occupy non-overlapping sites within
each moir\'e period {[}see the Fig.~\ref{fig:model_and_overview}(c){]},
minimizing interlayer repulsion and preserving lattice symmetry. If
the filling factors deviate from $(\rho_{\uparrow}=1/5,\rho_{\downarrow}=4/5)$
for $\theta=36.87^{\circ}$, the remaining particles or vacant sites
are likely to break the moir\'e lattice symmetry. Similarly, the system
with other twist angle {[}e.g. the $(\theta=22.62^{\circ},\rho_{\uparrow}=2/5,\rho_{\downarrow}=3/5)$
case shown in the lower left plot of Fig.~\ref{fig:model_and_overview}(d){]}
also hosts this kind of ``interlocked'' MI with moir\'e lattice symmetry.
By contrast, for a SF phase with moir\'e lattice symmetry, the filling
factor requirements are less restrictive because the number of atoms
per site is not constrained to integer values, allowing particles
to distribute themselves in accordance with the moir\'e lattice symmetry.
We also notice that the interplay between interlayer repulsion and
filling imbalance gives rise to the difference in the critical $J/U(g_{\sigma\sigma})$
shown in Figs.~\ref{fig:balanced_and_imbalanced_filling}(a) and
(b): in the imbalanced case ($\rho_{\uparrow}=1/5$), the larger interparticle
distance on the $\uparrow$-layer requires a higher $J/U(g_{\sigma\sigma})$
for coherence and transition into the SF phase compared to the balanced
case ($\rho_{\downarrow}=4/5$).}

Although the dynamical moir\'e system can manifest SF or MI phase similar
to the ones found in conventional bosonic moir\'e systems \cite{meng_Atomic_2023},
there are fundamental differences. Instead of being coupled by interlayer
tunneling, the atoms on different layers are coupled by the interlayer
density-density interactions, { which preserve the independent U(1)
symmetries in each layer.} Consequently, the total particle number
of each layer is conserved and can be different. These fundamental
differences thus naturally indicate much richer physics with no counterparts
in conventional bosonic moir\'e systems can emerge, as we shall discuss
below.

\subsection{Beyond the conventional moir\'e physics}

\begin{figure}[H]
\begin{centering}
\includegraphics[width=3.1in]{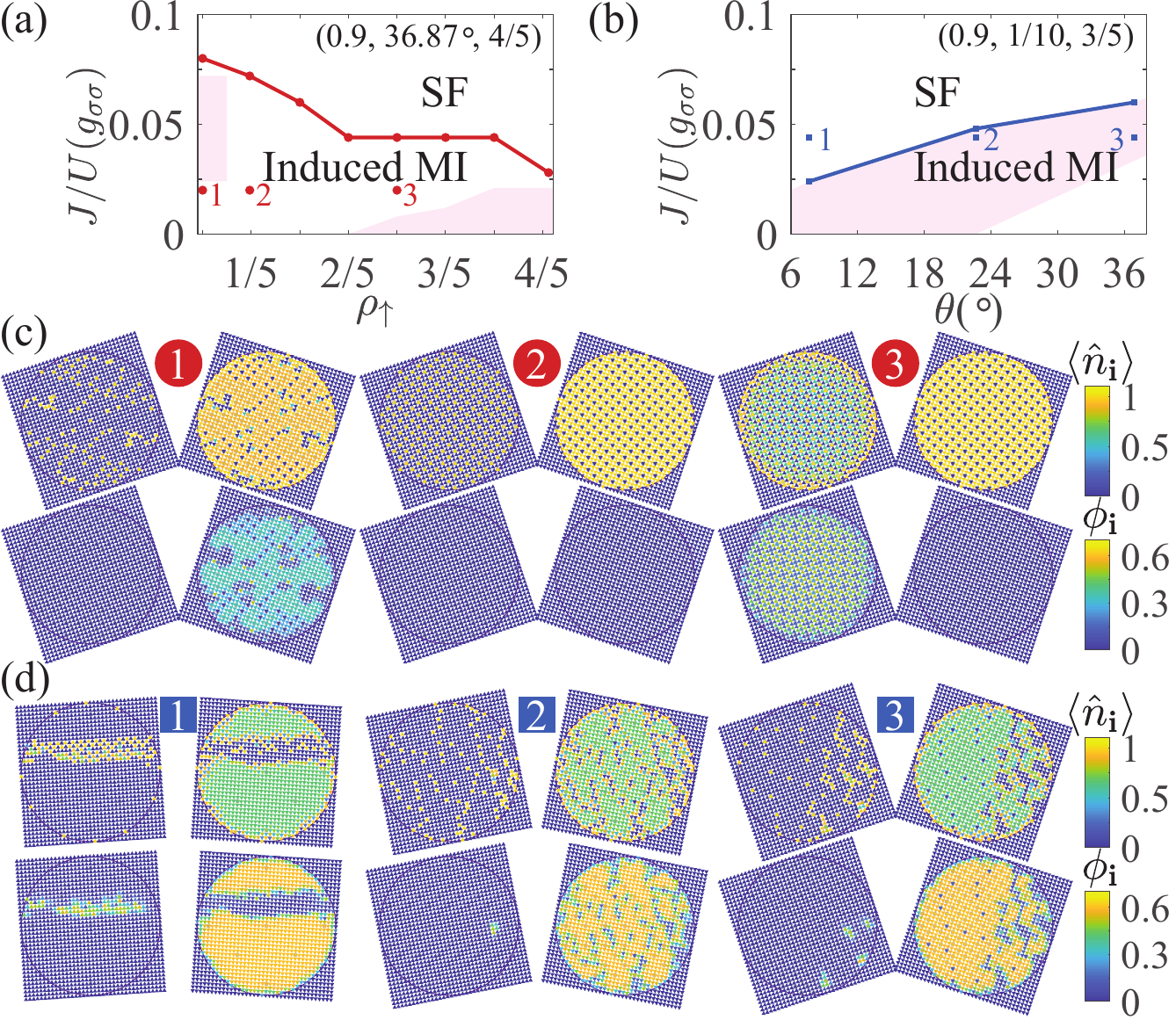} 
\par\end{centering}
\caption{(a) Phase diagram with respect to $\rho_{\uparrow}$ and $J/U(g_{\sigma\sigma})$
with other parameters fixed at $g_{\uparrow\downarrow}/g_{\sigma\sigma}=0.9$,
$\theta=36.87^{\circ}$, and $\rho_{\downarrow}=4/5$. (b) Phase diagram
with respect to the twist angle $\theta$ and $J/U(g_{\sigma\sigma})$
with other parameters fixed at $g_{\uparrow\downarrow}/g_{\sigma\sigma}=0.9$,
$\rho_{\uparrow}=1/10$, and $\rho_{\downarrow}=3/5$. {{} The magenta
regions of the phase diagrams contain the BG phase.} (c) and (d)
Real-space particle density and superfluid order parameter distributions
of the system that correspond to the red dots and the blue squares
in (a) and (b), respectively. }
\label{fig:filling_diff_twist_angle_physics} 
\end{figure}

Since the total particle number of each layer is conserved separately,
the filling difference between the two layers in fact provides an
extra tuning knob of new physics compared with the conventional bosonic
moir\'e systems, as showcased in Fig.~\ref{fig:balanced_and_imbalanced_filling}.
In Fig.~\ref{fig:filling_diff_twist_angle_physics}(a), we systematically
investigate the influences of the filling difference by keeping the
filling factor of the $\downarrow$-layer fixed at $\rho_{\downarrow}=4/5$
and changing the filling factor of the $\uparrow$-layer $\rho_{\uparrow}$
from $1/10$ to $4/5$. As we can see from the {Fig.~\ref{fig:filling_diff_twist_angle_physics}(c1)},
when the filling of the $\uparrow$-layer is small ($\rho_{\uparrow}=1/10$),
the $\downarrow$-layer is in a SF phase and it imposes strong influences
on atoms on the $\uparrow$-layer by localizing them and making them
in induced MI phase despite $\rho_{\uparrow}=1/10$ being non-integer.
As the $\rho_{\uparrow}$ gradually increases {[}see {Figs.~\ref{fig:filling_diff_twist_angle_physics}(c2)
and (c3)}{]}, the $\uparrow$-layer also imposes an enhanced localization
effects on the atoms on the $\downarrow$-layer by driving a SF-MI
transition on this layer. We also observe that the average localization
effect experienced by each atom in the $\uparrow$-layer decreases,
making the $\uparrow$-layer transition into a SF phase at relatively
high filling with $\rho_{\uparrow}=1/2$. This phenomenon is reminiscent
of what happens in high-temperature superconductors, where doping
can lead to significant changes in the system's transport properties
\cite{Zaanen_Nature_2015}.

Similar to the conventional moir\'e system, the twist angle also has
a significant impact here, as clearly shown in Fig.~\ref{fig:filling_diff_twist_angle_physics}(b).
For the case at the filling $(\rho_{\uparrow}=1/10,\rho_{\downarrow}=3/5)$,
in the intermediate hopping regime, we see that at a relatively small
twist angle $\theta=7.63^{\circ}$, the system is in the SF$_{\uparrow}$-SF$_{\downarrow}$
phase. As the twist angle increases to $\theta=22.62^{\circ}$, the
atoms in the $\uparrow$-layer are localized due to the interaction
from the $\downarrow$-layer and enter the induced MI phase, forming
the phase MI$_{\uparrow}$-SF$_{\downarrow}$ {[}see {Fig.~\ref{fig:filling_diff_twist_angle_physics}(d2)}{]}.
Similar situation is also found at even larger twist angle, for instance,
$\theta=36.87^{\circ}$ {[}see {Fig.~\ref{fig:filling_diff_twist_angle_physics}(d3)}{]}.

Interestingly, we notice that density distributions in the {Figs.~\ref{fig:filling_diff_twist_angle_physics}(d2)
and (d3)} assume rather random pattern. In these cases, the atoms
on $\downarrow$-layer in fact provide a ``disordered'' effective
potential to the atoms lying on the $\uparrow$-layer. Thus, one naturally
expects that this ``disordered'' effective potential quenches hopping
of the $\uparrow$-atoms and makes them localized, which can be interpreted
as the interlayer interaction induced ``self-localization'' between
the atoms on the two layers. {The presence of the ``disordered''
effective potential is associated with phases with the spontaneous
breaking of $C_{4}$ rotational symmetry of the system {[}see Fig.~\ref{fig:filling_diff_twist_angle_physics}(c1)
and Fig.~\ref{fig:filling_diff_twist_angle_physics}(d){]}, indicating
that there are four degenerate ground states (see the Appendix~\ref{sec:C4}
for the degenerate ground states).}

Among the states with irregular density pattern, such as the {Fig.~\ref{fig:filling_diff_twist_angle_physics}(d3)}
and the last plot of Fig. \ref{fig:model_and_overview}(d), there
exist states with disconnected superfluid islands, exhibiting the
characteristics of {BG} (see the Appendix~\ref{sec:BG} for exact identification). Thus, this system supports BG without extrinsic
disorder {(see the magenta region corresponding to the BG phase in
the phase diagrams shown in Figs.~2 and 3)}, in contrast to previous
studies with either (quasi)disorder or quasicrystalline lattices \cite{fisher1989boson,fallani2007ultracold,pasienski2010disordered,soyler2011phase,gadway2011glassy,derrico2014observation,meldgin2016probing,Johnstone_2021,ciardi2023quasicrystalline}.
{BG phase in our system is characterized by the presence of isolated,
non-percolated superfluid islands, and usually surrounded by MI regions.
Therefore, in the phase diagrams, the BG phase is located within the
broader ``induced MI'' region, except the filling-angle-match case
{[}see the Fig.~\ref{fig:balanced_and_imbalanced_filling} (b){]}
whose density distributions with the moir\'e lattice symmetry would
not provide an effective disordered potential.} {Consistent with
previous studies, we have found many metastable configurations associated
with the BG phase in our system (see the Appendix~\ref{sec:C4} for
the metastable configurations).} As the presence of BG at ground
state usually implies many-body localization (MBL) at excited states
\cite{RMP2019MBL}, this system also probably hosts (quasi-)MBL without
extrinsic disorder \cite{schiulaz_Dynamics_2015,yao2016quasiMBL,smith_DisorderFree_2017,karpov_DisorderFree_2021}.
Note that here either the BG or the possible quasi-MBL are found at
commensurate twist angles, i.e., the system has translational invariance
in the bulk, in contrast to previously investigated localization at
incommensurate angles \cite{huang_Moir_2019,paul2023particle,madronero2024localized}
or the interacting 2D Aubry-Andr\'e model with quasi-periodicity \cite{szabo_Mixed_2020,strkalj_Coexistence_2022}.
The possible quasi-MBL here may be explained by a mechanism similar
to the exponentially slow relaxation predicted in certain translation-invariant
systems \cite{schiulaz_Dynamics_2015,yao2016quasiMBL}. Further exploration
of the relation between the localization found here with MBL and the
generalization to incommensurate twist angles will be an intriguing
subject for future study. 

In light of the tremendous experimental advancements in ultracold
atomic physics, we believe that the rich physics in this dynamical
moir\'e system can be readily observed using current experimental setups
\cite{meng_Atomic_2023}. More specifically, the $J/U(g_{\sigma\sigma})$
can be tuned by adjusting the depth of optical lattice potential.
The interlayer interaction $U_{\mathbf{i}_{\uparrow}\mathbf{j}_{\downarrow}}(g_{\uparrow\downarrow},\theta)$
can be altered by either tuning the vertical displacement between
the two spin-dependent lattices, or by utilizing the Feshbach resonance
\footnote{The Feshbach resonance technique enables the direct tuning of $g_{\uparrow\downarrow}/g_{\sigma\sigma}$
in the phase diagrams. For the interlayer interaction $U_{\mathbf{i}_{\uparrow}\mathbf{j}_{\downarrow}}\equiv\frac{g_{\uparrow\downarrow}}{2}\int d^{3}\mathbf{r}\left|w_{\mathbf{i}_{\uparrow}}(\mathbf{r})\right|^{2}\left|w_{\mathbf{j}_{\downarrow}}(\mathbf{r})\right|^{2}$,
displacing the spin-dependent lattices in out-of-plane direction reduces
the overlap between Wannier functions $w_{{\uparrow/\downarrow}}(\mathbf{r})$,
whose effect is physically equivalent to the change of $g_{\uparrow\downarrow}/g_{\sigma\sigma}$}. The control over the twist angle can be realized by experimental
techniques developed recently for the twisted bilayer optical lattices
\cite{meng_Atomic_2023}. The exotic SF, MI phases, and various self-localization
behavior can be directly observed via the quantum gas microscope \cite{Gross_Nat_Phys_2021}.

\section{Conclusions}

{ We have proposed a dynamical moir\'e system induced purely by interlayer
particle interaction in twisted bilayer optical lattices,} which
hosts a wide range of unique physical phenomena. 
both the dynamical moir\'e system itself and its experimental realization
is even simpler than the conventional bosonic moir\'e system realized
in experiments recently \cite{meng_Atomic_2023}, we believe our work
will stimulate further theoretical research on this type of dynamical
moir\'e system, especially the possible intrinsic connection between
the self-localization found here and quasi-MBL, generalization to
fermionic systems, as well as experimental effort in searching for
these exotic quantum phases. {In addition, while not present in our
purely repulsive interlayer system, other exotic quantum phases like
pair superfluid and dimerized phases could exist in bilayer systems
with attractive interlayer interactions \cite{psf1,psf2}, warranting
further study.} 

\begin{acknowledgements}
This work is supported by NKRDPC (Grant No.~2022YFA1405304), NSFC
(Grant Nos.~12004118 and 12275089), Guangdong Basic and Applied Basic
Research Foundation (Grants Nos.~2020A1515110228, 2021A1515010212
and 2023A1515012800), Guangdong Provincial Key Laboratory (Grant No.~2020B1212060066),
and START grant of South China Normal University. 
\end{acknowledgements}

\appendix

\section{Model Hamiltonian and Gutzwiller method}\label{sec:System_model}


The low physics of this system can be described by a Bose-Hubbard
type model within the lowest band approximation. To derive the model
Hamiltonian, we start with the following many-body Hamiltonian 
\begin{align}
\hat{H}= & \sum_{\sigma=\uparrow,\downarrow}\int d^{3}\mathbf{r}\hat{\psi}_{\sigma}^{\dagger}(\mathbf{r})\left(\frac{-\hbar^{2}}{2\mu}\nabla^{2}+V(\mathrm{\boldsymbol{r}})\right)\hat{\psi_{\sigma}}(\mathbf{r})\label{eq:MBH}\\
 & +\sum_{\sigma=\uparrow,\downarrow}\int d^{3}\mathbf{r}\int d^{3}\mathbf{r}^{\prime}\left[\hat{\psi}_{\sigma}^{\dagger}(\mathbf{r})\hat{\psi}_{\sigma}^{\dagger}(\mathbf{r^{\prime}})\frac{V_{\sigma\sigma}(\mathbf{r}-\mathbf{r^{\prime}})}{2}\right.\nonumber \\
 & \left.\hat{\psi}_{\sigma}(\mathbf{r^{\prime}})\hat{\psi}_{\sigma}(\mathbf{r})\right]\nonumber \\
 & +\int d^{3}\mathbf{r}\int d^{3}\boldsymbol{\mathbf{r}}^{\prime}\hat{\psi}_{\uparrow}^{\dagger}(\mathbf{r})\hat{\psi}_{\downarrow}^{\dagger}(\mathbf{r^{\prime}})V_{\uparrow\downarrow}(\mathbf{r}-\mathbf{r^{\prime}})\hat{\psi}_{\downarrow}(\mathbf{r^{\prime}})\hat{\psi}_{\uparrow}(\mathbf{r}),\nonumber 
\end{align}
where $\sigma=\uparrow,\downarrow$ is the index of the atom species
and lattice layer, $V_{\sigma\sigma}(\mathbf{r}-\mathbf{r^{\prime}})\left[V_{\uparrow\downarrow}(\mathbf{r}-\mathbf{r^{\prime}})\right]$
represents the interaction between two intra(inter)-species atoms,
$V(\mathrm{\boldsymbol{r}})$ represents the lattice potential, $\mu$
is the mass of an atom, and the operators $\hat{\psi_{\sigma}}^{\dagger}(\mathbf{r})$
and $\hat{\psi_{\sigma}}(\mathbf{r})$ respectively create and annihilate
an atom at position $\boldsymbol{\mathbf{r}}$ of the lattice layer
$\sigma$. In the context of ultracold atomic gases, where the interactions
are modeled by effective two-particle interactions for suciently dilute
gases, a choice suitable for most theoretical models is the delta
contact potential between two atoms of mass $\mu$ at positions $\mathbf{r}$
and $\mathbf{r^{\prime}}$ 
\begin{align}
V_{\sigma\sigma}(\mathbf{r}-\mathbf{r^{\prime}}) & =\frac{4\pi\hbar^{2}a_{\sigma\sigma}}{\mu}\delta(\mathbf{r}-\mathbf{r^{\prime}}),\nonumber \\
V_{\uparrow\downarrow}(\mathbf{r}-\mathbf{r^{\prime}}) & =\frac{4\pi\hbar^{2}a_{\uparrow\downarrow}}{\mu}\delta(\mathbf{r}-\mathbf{r^{\prime}}),
\end{align}
with $a_{\sigma\sigma}$ ($a_{\uparrow\downarrow}$) being the intra(inter)-species
$s$-wave scattering length. The first term of the Hamiltonian (\ref{eq:MBH})
is the single particle term, and the second term describes the intra-species
interaction of atoms in the same lattice layer, and the third term
describes the inter-species interaction between atoms of different
lattice layer.

Within the second quantization formalism, we can define the creation
operator for the lowest band Wannier orbital $w_{\mathbf{i}_{\sigma}}(\mathbf{r})=\langle\mathbf{r}\mid\mathbf{i}_{\sigma}\rangle$
at site $\mathbf{i}$ of $\sigma$-layer, 
\begin{equation}
\hat{b}_{\mathbf{i}_{\sigma}}^{\dagger}=\int d^{3}\mathbf{r}w_{\mathbf{i}_{\sigma}}(\mathbf{r})\hat{\psi}_{\sigma}^{\dagger}(\mathbf{r}).
\end{equation}

For a relatively shallow optical lattice, the single particle Hamiltonian
within the tight-binding approximation restricted to lowest band reads
\begin{align}
H_{\mathrm{s.p.},\sigma}= & \int d^{3}\mathbf{r}\hat{\psi}_{\sigma}^{\dagger}(\mathbf{r})\left(\frac{-\hbar^{2}}{2\mu}\nabla^{2}+V(\mathrm{\boldsymbol{r}})\right)\hat{\psi_{\sigma}}(\mathbf{r})\\
= & \sum_{\mathbf{i}_{\sigma},\mathbf{j}_{\sigma}}\int d^{3}\mathbf{r}w_{\mathbf{i}_{\sigma}}^{*}(\mathbf{r})\hat{b}_{\mathbf{i}_{\sigma}}^{\dagger}\left(\frac{-\hbar^{2}}{2\mu}\nabla^{2}+V(\mathrm{\boldsymbol{r}})\right)w_{\mathbf{j}_{\sigma}}(\mathbf{r})\hat{b}_{\mathbf{j}_{\sigma}}\nonumber \\
= & \sum_{\mathbf{i}_{\sigma}}\epsilon_{\mathbf{i}_{\sigma}}\hat{b}_{\mathbf{i}_{\sigma}}^{\dagger}\hat{b}_{\mathbf{i}_{\sigma}}-J\sum_{\left\langle \mathbf{i}_{\sigma},\mathbf{j}_{\sigma}\right\rangle }\hat{b}_{\mathbf{i}_{\sigma}}^{\dagger}\hat{b}_{\mathbf{j}_{\sigma}},\nonumber 
\end{align}
where $\langle\cdots\rangle$ denotes the lattice sites within the
nearest-neighbor lattice sites, with 
\begin{equation}
\epsilon_{\mathbf{i}_{\sigma}}\equiv\int d^{3}\mathbf{r}w_{\mathbf{i}_{\sigma}}^{*}(\mathbf{r})[\frac{-\hbar^{2}}{2\mu}\nabla^{2}+V(\mathrm{\boldsymbol{r}})]w_{\mathbf{i}_{\sigma}}(\mathbf{r}),
\end{equation}
and 
\begin{equation}
J\equiv-\int d^{3}\mathbf{r}w_{\mathbf{i}_{\sigma}}^{*}(\mathbf{r})[\frac{-\hbar^{2}}{2\mu}\nabla^{2}+V(\mathrm{\boldsymbol{r}})]w_{\mathbf{j}_{\sigma}}(\mathbf{r}).
\end{equation}

For a sufficiently dilute, weakly interacting atomic gas, the short
ranged interactions covering the nearest neighbor sites of single
component are taken into consideration. The interaction Hamiltonian
of single component reads 
\begin{align}
H_{\mathrm{intra},\sigma}= & \int d^{3}\mathbf{r}\int d^{3}\mathbf{r}^{\prime}\hat{\psi}_{\sigma}^{\dagger}(\mathbf{r})\hat{\psi}_{\sigma}^{\dagger}(\mathbf{r^{\prime}})\frac{V_{\sigma\sigma}(\mathbf{r}-\mathbf{r^{\prime}})}{2}\hat{\psi}_{\sigma}(\mathbf{r^{\prime}})\hat{\psi}_{\sigma}(\mathbf{r})\nonumber \\
= & \sum_{\mathbf{i}_{\sigma},\mathbf{j}_{\sigma},\mathbf{k}_{\sigma},\mathbf{l}_{\sigma}}\frac{1}{2}\int d^{3}\mathbf{r}\int d^{3}\mathbf{r}^{\prime}\left[w_{\mathbf{i}_{\sigma}}^{*}(\mathbf{r})w_{\mathbf{j}_{\sigma}}^{*}(\mathbf{r^{\prime}})\frac{4\pi\hbar^{2}a_{\sigma\sigma}}{\mu}\right.\nonumber \\
 & \left.\delta(\mathbf{r}-\mathbf{r^{\prime}})w_{\mathbf{k}_{\sigma}}(\mathbf{r^{\prime}})w_{\mathbf{l}_{\sigma}}(\mathbf{r})\hat{b}_{\mathbf{i}_{\sigma}}^{\dagger}\hat{b}_{\mathbf{j}_{\sigma}}^{\dagger}\hat{b}_{\mathbf{k}_{\sigma}}\hat{b}_{\mathbf{l}_{\sigma}}\right]\\
= & \sum_{\mathbf{i}_{\sigma}}\frac{U_{\mathbf{i}_{\sigma}\mathbf{i}_{\sigma}}}{2}\hat{b}_{\mathbf{i}_{\sigma}}^{\dagger}\hat{b}_{\mathbf{i}_{\sigma}}^{\dagger}\hat{b}_{\mathbf{i}_{\sigma}}\hat{b}_{\mathbf{i}_{\sigma}}+\sum_{\left\langle \mathbf{i}_{\sigma},\mathbf{j}_{\sigma}\right\rangle }\frac{U_{\mathbf{i}_{\sigma}\mathbf{j}_{\sigma}}}{2}\hat{b}_{\mathbf{i}_{\sigma}}^{\dagger}\hat{b}_{\mathbf{i}_{\sigma}}\hat{b}_{\mathbf{j}_{\sigma}}^{\dagger}\hat{b}_{\mathbf{j}_{\sigma}}\nonumber 
\end{align}
where 
\begin{equation}
U_{\mathbf{i}_{\sigma}\mathbf{i}_{\sigma}}\equiv\frac{g_{\sigma\sigma}}{2}\int d^{3}\mathbf{r}\left|w_{\mathbf{i}_{\sigma}}(\mathbf{r})\right|^{4},
\end{equation}
and 
\begin{equation}
U_{\mathbf{i}_{\sigma}\mathbf{j}_{\sigma}}\equiv\frac{g_{\sigma\sigma}}{2}\int d^{3}\mathbf{r}\left|w_{\mathbf{i}_{\sigma}}(\mathbf{r})\right|^{2}\left|w_{\mathbf{j}_{\sigma}}(\mathbf{r})\right|^{2}.
\end{equation}
Here $g_{\sigma\sigma}=4\pi\hbar^{2}a_{\sigma\sigma}/\mu$ denotes
the intra-species contact interaction strength.

Accordingly, the interaction Hamiltonian between different components
takes the form that 
\begin{equation}
H_{\mathrm{inter}}=\sum_{\langle\mathbf{i}_{\uparrow},\mathbf{j}_{\downarrow}\rangle}U_{\mathbf{i}_{\uparrow}\mathbf{j}_{\downarrow}}\hat{n}_{\mathbf{i}_{\uparrow}}\hat{n}_{\mathbf{j}_{\downarrow}},
\end{equation}
where 
\begin{equation}
U_{\mathbf{i}_{\uparrow}\mathbf{j}_{\downarrow}}\equiv\frac{g_{\uparrow\downarrow}}{2}\int d^{3}\mathbf{r}\left|w_{\mathbf{i}_{\uparrow}}(\mathbf{r})\right|^{2}\left|w_{\mathbf{j}_{\downarrow}}(\mathbf{r})\right|^{2},
\end{equation}
with $g_{\uparrow\downarrow}=4\pi\hbar^{2}a_{\uparrow\downarrow}/\mu$
being the inter-species contact interaction strength.

Considering the 2D case, we approximate the Wannier function \textbf{$w_{\mathbf{i}_{\sigma}}(\mathbf{r})$}
as a harmonic oscillator wave function of the lowest energy level,
therefore, the Wannier function (at site $\mathbf{0}$ for instance)
in the 2D plane reads 
\begin{equation}
w(x,y)=\sqrt{2\kappa}e^{-\pi\kappa(x^{2}+y^{2})},\label{eq:Wannier=00003D000020function-1}
\end{equation}
where $\kappa\equiv\sqrt{\mu V_{0}/(2\hbar^{2}d^{2})}$ with $V_{0}$
being the lattice depth and $d$ being the lattice constant. As a
result, the interlayer interaction strength can be expressed as 
\begin{equation}
U_{\mathbf{i}_{\uparrow}\mathbf{j}_{\downarrow}}=\frac{g_{\uparrow\downarrow}}{2}\kappa e^{-\pi\kappa\left(\Delta_{x}^{2}+\Delta_{y}^{2}\right)},
\end{equation}
where $\Delta_{x}^{2}+\Delta_{y}^{2}\equiv|\mathbf{r_{\mathbf{i}_{\uparrow}}}-\mathbf{r_{\mathbf{j}_{\downarrow}}}|^{2}$,
which depends on the twisted angle $\theta$.

Similarly, the strength of the interactions takes the form that 
\begin{equation}
\begin{cases}
U_{\mathbf{i}_{\sigma}\mathbf{i}_{\sigma}}= & \frac{g_{\sigma\sigma}}{2}\kappa\\
U_{\mathbf{i}_{\sigma}\mathbf{j}_{\sigma}}= & \frac{g_{\sigma\sigma}}{2}\kappa e^{-\pi\kappa d^{2}}\\
U_{\mathbf{i}_{\uparrow}\mathbf{j}_{\downarrow}}= & \frac{g_{\uparrow\downarrow}}{2}\kappa e^{-\pi\kappa\left(\Delta_{x}^{2}+\Delta_{y}^{2}\right)}
\end{cases}.
\end{equation}

In summary, the Hamiltonian of the system reads 
\begin{align}
\hat{H}= & \sum_{\sigma=\uparrow,\downarrow}\left[-J\sum_{\langle\mathbf{i}_{\sigma},\mathbf{j}_{\sigma}\rangle}\hat{b}_{\mathbf{i_{\sigma}}}^{\dagger}\hat{b}_{\mathbf{j_{\sigma}}}+\sum_{\mathbf{i}_{\sigma}}\frac{U(g_{\sigma\sigma})}{2}\hat{n}_{\mathbf{i}_{\sigma}}(\hat{n}_{\mathbf{i}_{\sigma}}-1)\right]\nonumber \\
 & +\sum_{\sigma=\uparrow,\downarrow}\sum_{\langle\mathbf{i}_{\sigma},\mathbf{j}_{\sigma}\rangle}\frac{U_{\mathbf{i}_{\sigma}\mathbf{j}_{\sigma}}(g_{\sigma\sigma})}{2}\hat{n}_{\mathbf{i}_{\sigma}}\hat{n}_{\mathbf{j}_{\sigma}}\nonumber \\
 & +\sum_{\langle\mathbf{i}_{\uparrow},\mathbf{j}_{\downarrow}\rangle}U_{\mathbf{i}_{\uparrow}\mathbf{j}_{\downarrow}}(g_{\uparrow\downarrow},\theta)\hat{n}_{\mathbf{i}_{\uparrow}}\hat{n}_{\mathbf{j}_{\downarrow}}.\label{eq:Hamiltonian-1}
\end{align}
where $\hat{n}_{\mathbf{i}_{\sigma}}\equiv\hat{b}_{\mathbf{i}_{\sigma}}^{\dagger}\hat{b}_{\mathbf{i}_{\sigma}}$
is the particle number operator that counts the number of atoms of
species $\sigma$ on site $\mathbf{i}$ of the $\sigma$-layer and
$U\equiv U_{\mathbf{i}_{\sigma}\mathbf{i}_{\sigma}}$.

We investigate the physics of this system by calculating its ground
state via the bosonic Gutzwiller variational approach \cite{Krauth_PRB_1992,Jaksch_PRL_1998,Lanata_PRB_2012}.
Within the framework of mean-field theory, we use the bosonic Gutzwiller
variational approach to investigate the ground state properties of
the system, with the variational ground state assumed to be the site-factorized
form 
\begin{equation}
|\mathrm{GW}\rangle=|\phi_{1_{\sigma}}\rangle_{1_{\sigma}}\otimes\ldots\otimes|\phi_{N_{\mathrm{lat},\sigma}}\rangle_{N_{\mathrm{lat},\sigma}}.
\end{equation}
Here, $N_{\mathrm{lat},\sigma}$ is the total number of the monolayer
lattice sites in $\sigma$-layer and $|\phi_{\mathbf{i}_{\sigma}}\rangle_{\mathbf{\mathbf{i}_{\sigma}}}=\sum_{n=0}^{\infty}c_{n}^{(\mathbf{\mathbf{i}_{\sigma})}}|n\rangle_{\mathbf{\mathbf{i}_{\sigma}}}$
is the local wave function at site $\mathbf{i}$ of $\sigma$-layer
with $|n\rangle_{\mathbf{i}_{\sigma}}$ being the corresponding local
occupation number state and $c_{n}^{(\mathbf{\mathbf{i}_{\sigma})}}$
being the variational parameter. The ground state is determined by
minimizing the total energy of the system within this variational
ansatz, i.e., 

\onecolumngrid
\begin{align}
E(\{c_{n}^{(\mathbf{\mathbf{i}_{\sigma})}}\})= & \langle\mathrm{GW}|\hat{H}|\mathrm{GW}\rangle\nonumber \\
= & \sum_{\mathbf{i}_{\sigma},\sigma=\uparrow,\downarrow}\left[\frac{U}{2}\frac{\sum_{n=0}^{\infty}\left|c_{n}^{(\mathbf{i}_{\sigma})}\right|^{2}n(n-1)}{\sum_{n=0}^{\infty}\left|c_{n}^{(\mathbf{i}_{\sigma})}\right|^{2}}+\sum_{\mathbf{j}_{\sigma}}\frac{U_{\mathbf{i}_{\sigma}\mathbf{j}_{\sigma}}}{2}\left(\frac{\sum_{n=0}^{\infty}\left|c_{n}^{(\mathbf{i}_{\sigma})}\right|^{2}n}{\sum_{n=0}^{\infty}\left|c_{n}^{(\mathbf{i}_{\sigma})}\right|^{2}}\right)\left(\frac{\sum_{n=0}^{\infty}\left|c_{n}^{(\mathbf{j}_{\sigma})}\right|^{2}n}{\sum_{n=0}^{\infty}\left|c_{n}^{(\mathbf{j}_{\sigma})}\right|^{2}}\right)\right.\nonumber \\
 & \left.-J\sum_{\mathbf{j}_{\sigma}}\left(\frac{\sum_{n=0}^{\infty}c_{n}^{*(\mathbf{i}_{\sigma})}c_{n+1}^{(\mathbf{i}_{\sigma})}\sqrt{n+1}}{\sum_{n=0}^{\infty}\left|c_{n}^{(\mathbf{i}_{\sigma})}\right|^{2}}\right)\left(\frac{\sum_{n=0}^{\infty}c_{n}^{(\mathbf{j}_{\sigma})}c_{n+1}^{*(\mathbf{j}_{\sigma})}\sqrt{n+1}}{\sum_{n=0}^{\infty}\left|c_{n}^{(\mathbf{j}_{\sigma})}\right|^{2}}\right)\right]\\
 & +\sum_{\langle\mathbf{i}_{\uparrow},\mathbf{j}_{\downarrow}\rangle}U_{\mathbf{i}_{\uparrow}\mathbf{j}_{\downarrow}}\left(\frac{\sum_{n=0}^{\infty}\left|c_{n}^{(\mathbf{i}_{\uparrow})}\right|^{2}n}{\sum_{n=0}^{\infty}\left|c_{n}^{(\mathbf{i}_{\uparrow})}\right|^{2}}\right)\left(\frac{\sum_{n=0}^{\infty}\left|c_{n}^{(\mathbf{j}_{\downarrow})}\right|^{2}n}{\sum_{n=0}^{\infty}\left|c_{n}^{(\mathbf{j}_{\downarrow})}\right|^{2}}\right).\nonumber 
\end{align}
\twocolumngrid

During the minimization process, we have tried different initial states
to obtain the global energy minimum.

\begin{figure*}[t]
\begin{centering}
\includegraphics[width=6.85in]{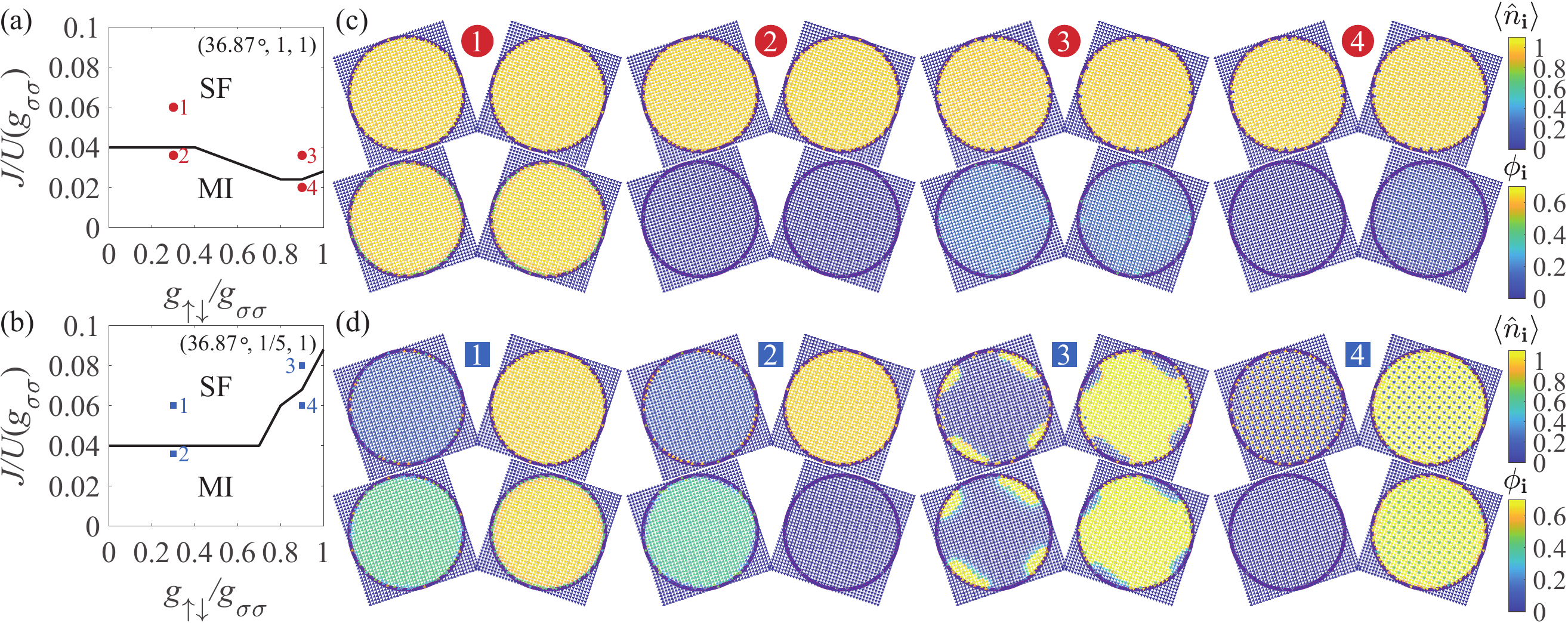}
\par\end{centering}
\caption{Phase diagrams and typical configurations of the MI-MI case and SF-MI
case. (a) Phase diagram with respect to the interlayer interaction
strength $g_{\uparrow\downarrow}/g_{\sigma\sigma}$ and the hopping
amplitude $J/U(g_{\sigma\sigma})$ at the filling with $\rho_{\uparrow}=1$
and $\rho_{\downarrow}=1$ (twist angle is fixed at $\theta=36.87^{\circ}$).
(b) Similar to (a), but at the filling with $\rho_{\uparrow}=1/5$
and $\rho_{\downarrow}=1$. (c) and (d) Real-space particle density
and superfluid order parameter distributions of the system that correspond
to the red dots and the blue squares in (a) and (b), respectively.}\label{fig:IntFilling}
\end{figure*}

\section{Phase diagrams for integer filling}\label{sec:integer_filling}

In the main text, we have focused on the SF-SF case, where both components
have non-integer fillings, rather than on the MI-MI case (both components
with integer fillings) or the SF-MI case (one component with a non-integer
filling and the other with an integer filling). This choice is due
to the large excitation gap caused by the strong interactions in the
integer-filling component, which hinders atom tunneling and the formation
of rich phases, unlike in the SF-SF case.

Here, we present the phase diagrams and typical configurations for
the MI-MI and SF-MI cases (see Fig~\ref{fig:IntFilling}). Similar
to the conventional Bose-Hubbard model, when the interlayer interaction
strength $g_{\uparrow\downarrow}/g_{\sigma\sigma}$ is small, the
components with integer filling favor the homogeneous superfluid phase
and Mott-insulator phase {[}see Figs~\ref{fig:IntFilling} (c1) and
(c2){]}, while the component with non-integer filling favors the homogeneous
superfluid phase {[}see Figs~\ref{fig:IntFilling} (d1) and (d2){]}.
In the MI-MI case, even when $g_{\uparrow\downarrow}/g_{\sigma\sigma}$
is large, the density distributions of both components remain almost
homogeneous due to the excitation gap {[}see Figs~\ref{fig:IntFilling}
(c3) and (c4){]}. Conversely, in the SF-MI case, a large $g_{\uparrow\downarrow}/g_{\sigma\sigma}$
drives the component with non-integer filling into an induced Mott-insulator
phase {[}see Fig~\ref{fig:IntFilling} (d4){]}. Additionally, the
phase boundary in Fig~\ref{fig:IntFilling} (a) reflects the system's
evolution as the interlayer interaction increases: from being dominated
by intralayer interactions, transitioning to a regime where interlayer
and intralayer interactions are in balanced competition, and finally
to a regime dominated by interlayer interactions. In the SF-MI scenario,
the high-filling component can create an effective potential that
localizes the low-filling component, thereby facilitating the formation
of the Mott-insulator phase as $g_{\uparrow\downarrow}/g_{\sigma\sigma}$
increases {[}see Fig~\ref{fig:IntFilling} (b){]}.

\begin{figure}[H]
\begin{centering}
\includegraphics[totalheight=2.5in]{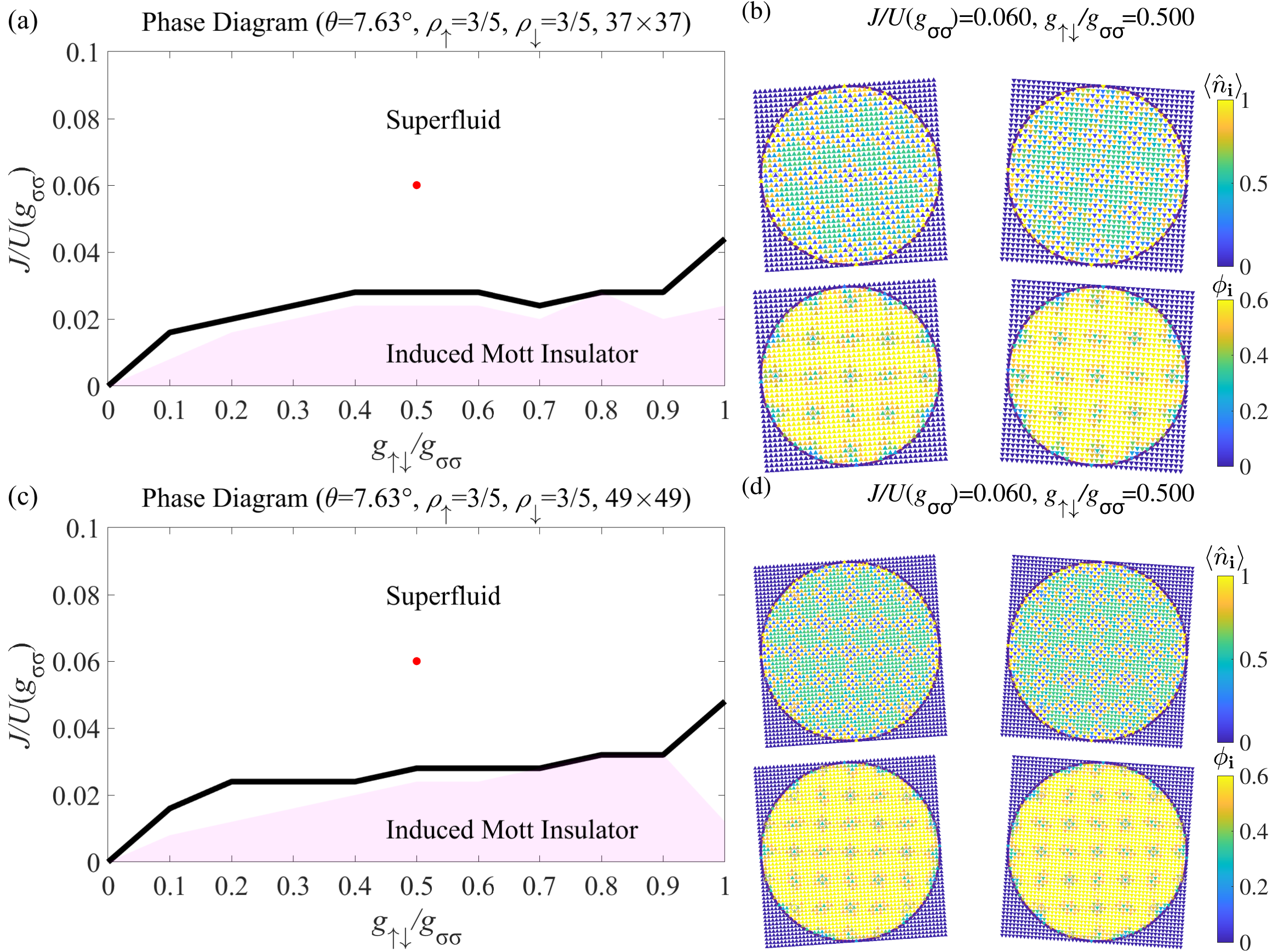}
\par\end{centering}
\caption{\textcolor{teal}{}(a) Phase diagram with respect to the interlayer
interaction strength $g_{\uparrow\downarrow}/g_{\sigma\sigma}$ and
the hopping amplitude $J/U(g_{\sigma\sigma})$ with $\rho_{\uparrow}=3/5$
and $\rho_{\downarrow}=3/5$ (the twist angle $\theta$ is fixed at
$\theta=7.63^{\circ}$ and the system size is $37\times37$). (b)
Real-space particle density and superfluid order parameter distributions
of the system that correspond to the red dot in (a). (c) Similar to
(a), but the system size is $49\times49$. (d) Real-space particle
density and superfluid order parameter distributions of the system
that correspond to the red dot in (c).}\label{fig:SystemSize}
\end{figure}

\begin{table*}[t]
\begin{centering}
\begin{tabular}{|c|c|c|c|}
\hline 
\multicolumn{4}{|c|}{Four degenerate ground states}\tabularnewline
\hline 
\includegraphics[scale=0.15]{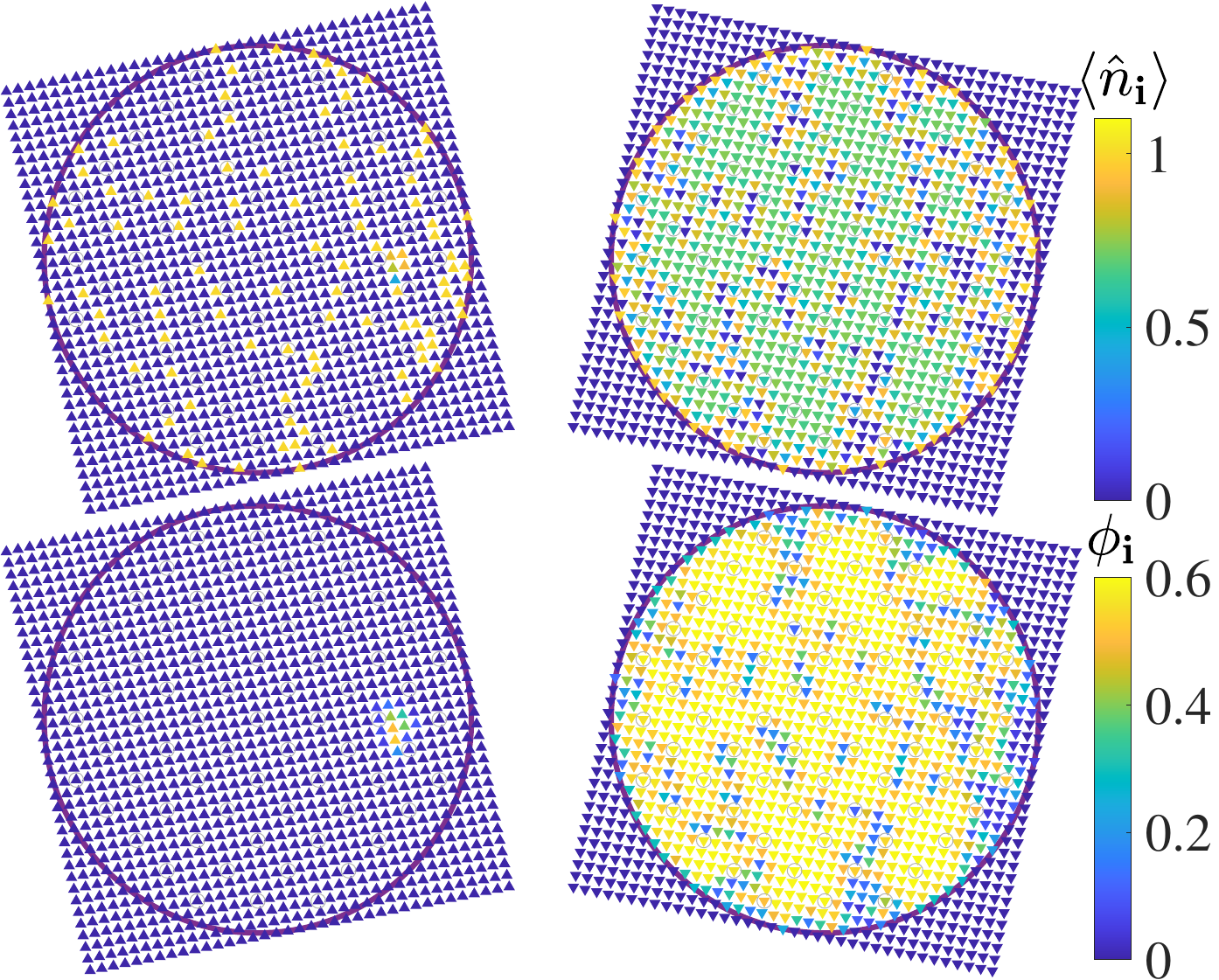} & \includegraphics[scale=0.15]{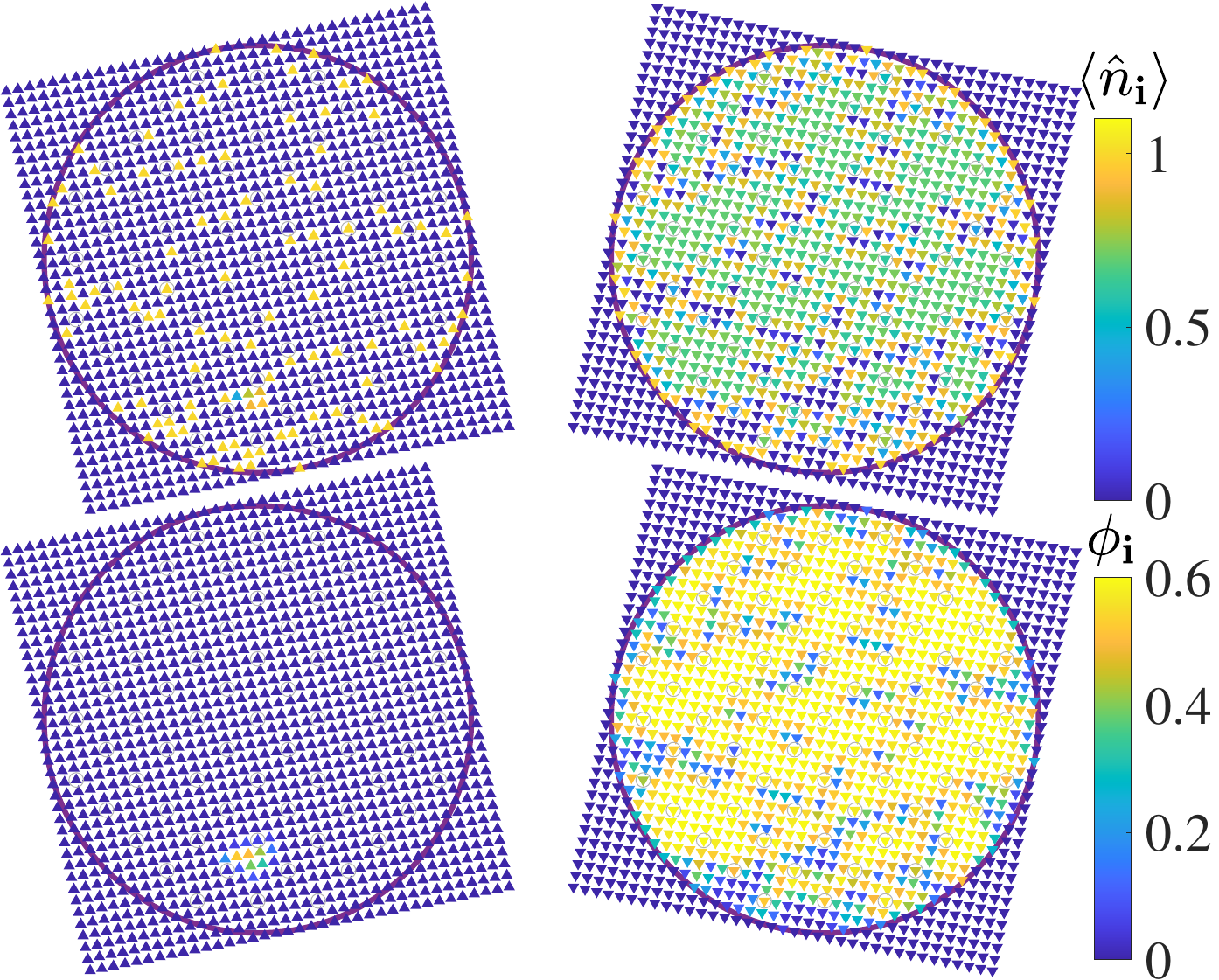} & \includegraphics[scale=0.15]{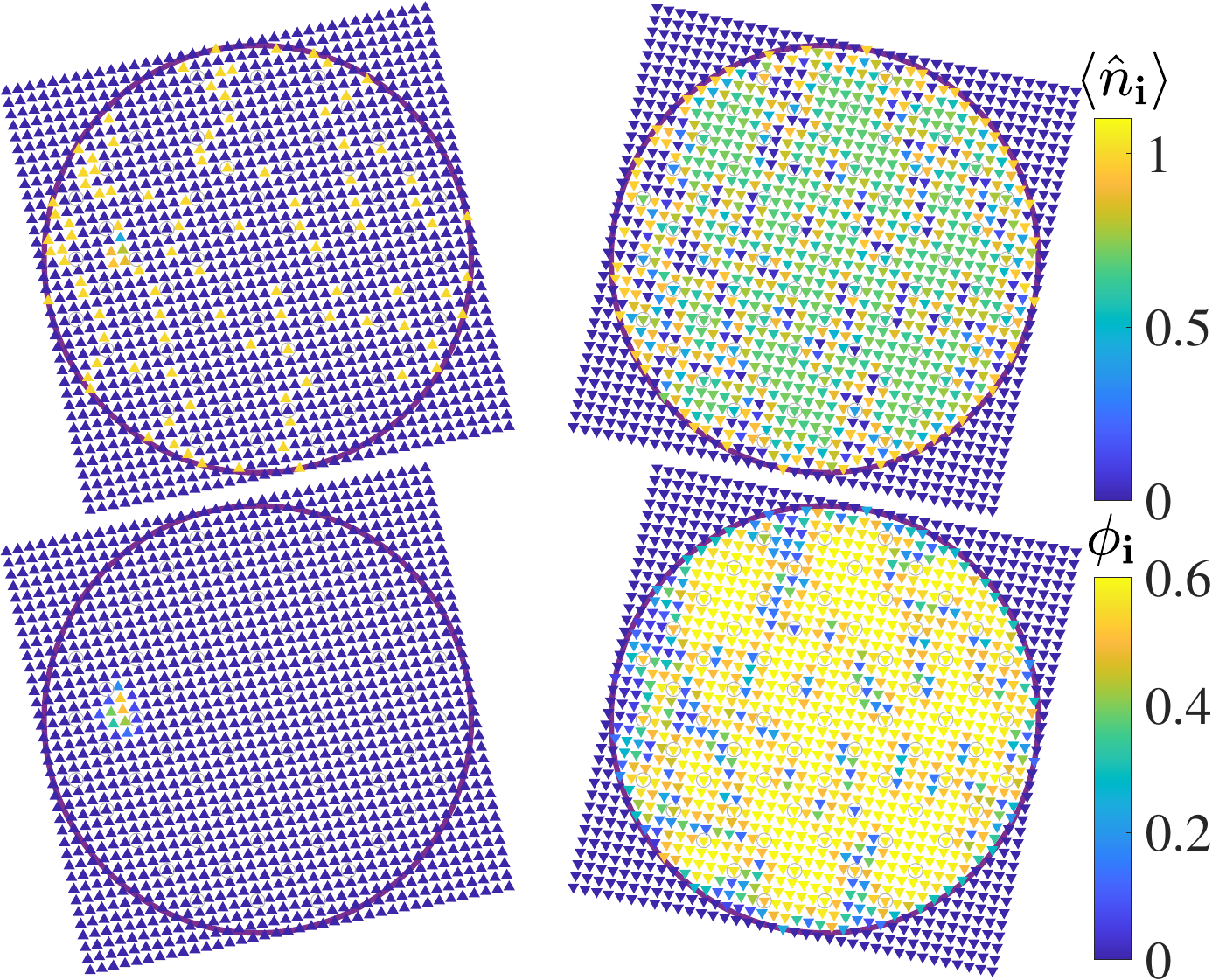} & \includegraphics[scale=0.15]{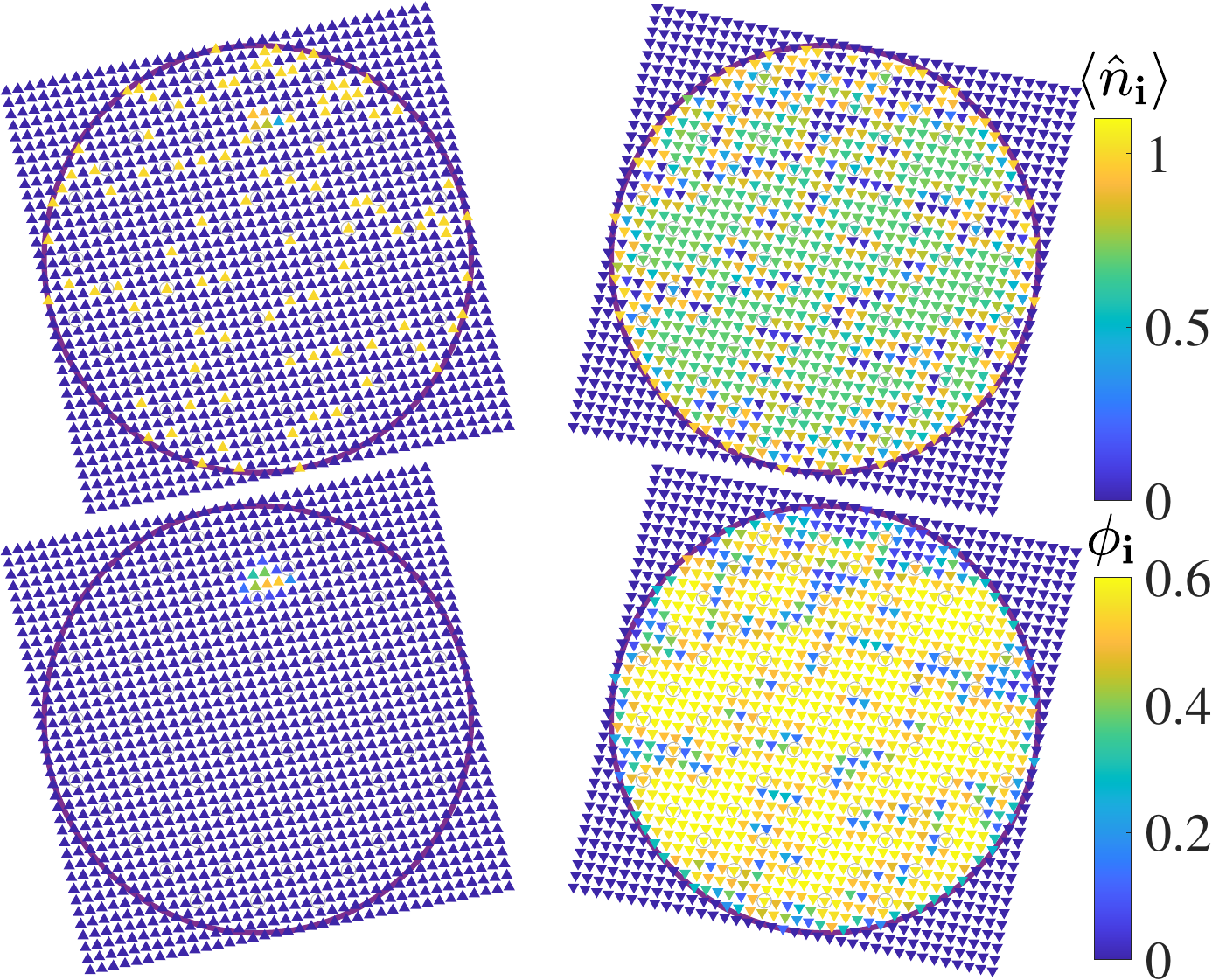}\tabularnewline
\hline 
$E=56.56U$ & $E=56.56U$ & $E=56.56U$ & $E=56.56U$\tabularnewline
\hline 
\multicolumn{4}{|c|}{Metastable States}\tabularnewline
\hline 
\includegraphics[scale=0.15]{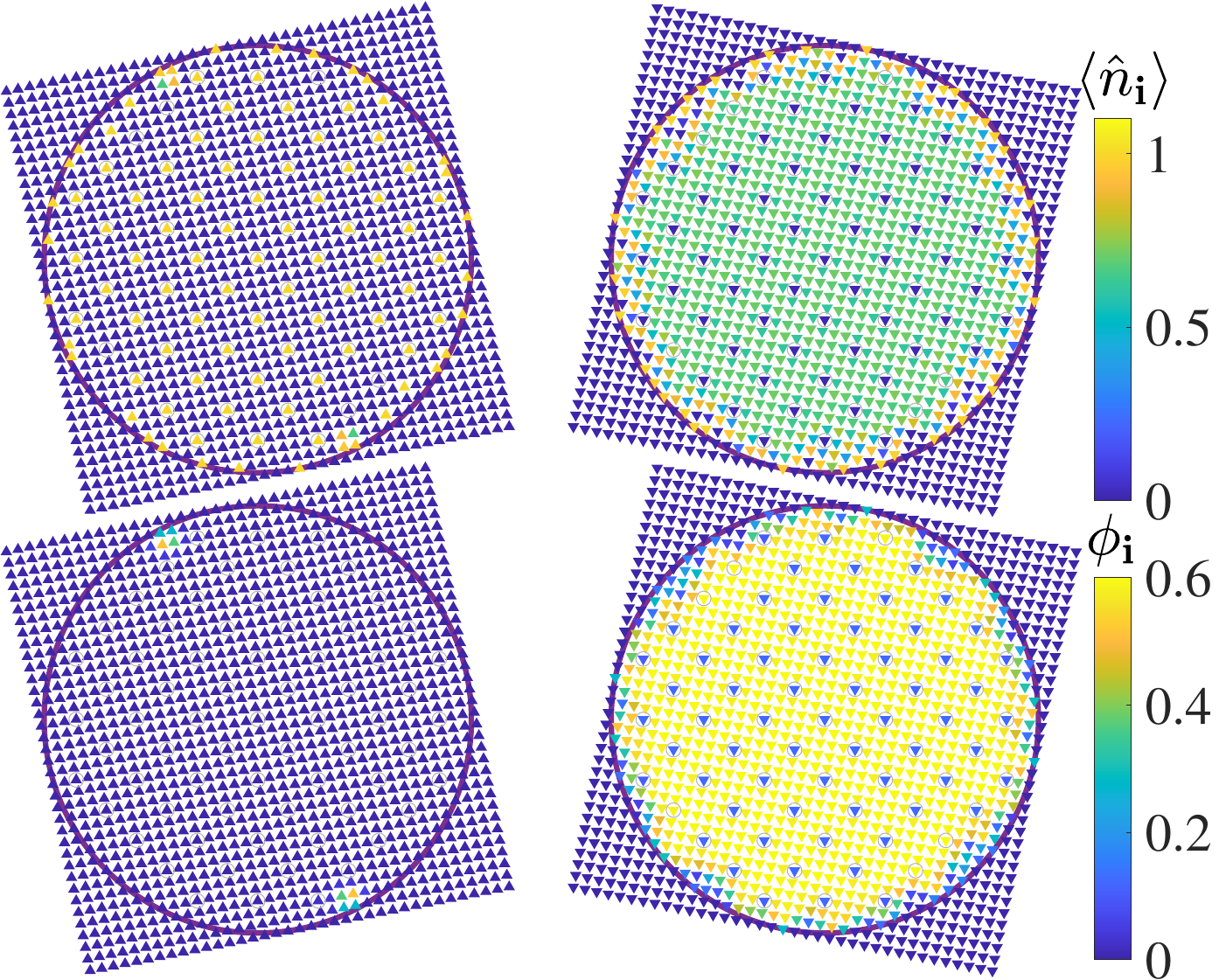} & \includegraphics[scale=0.15]{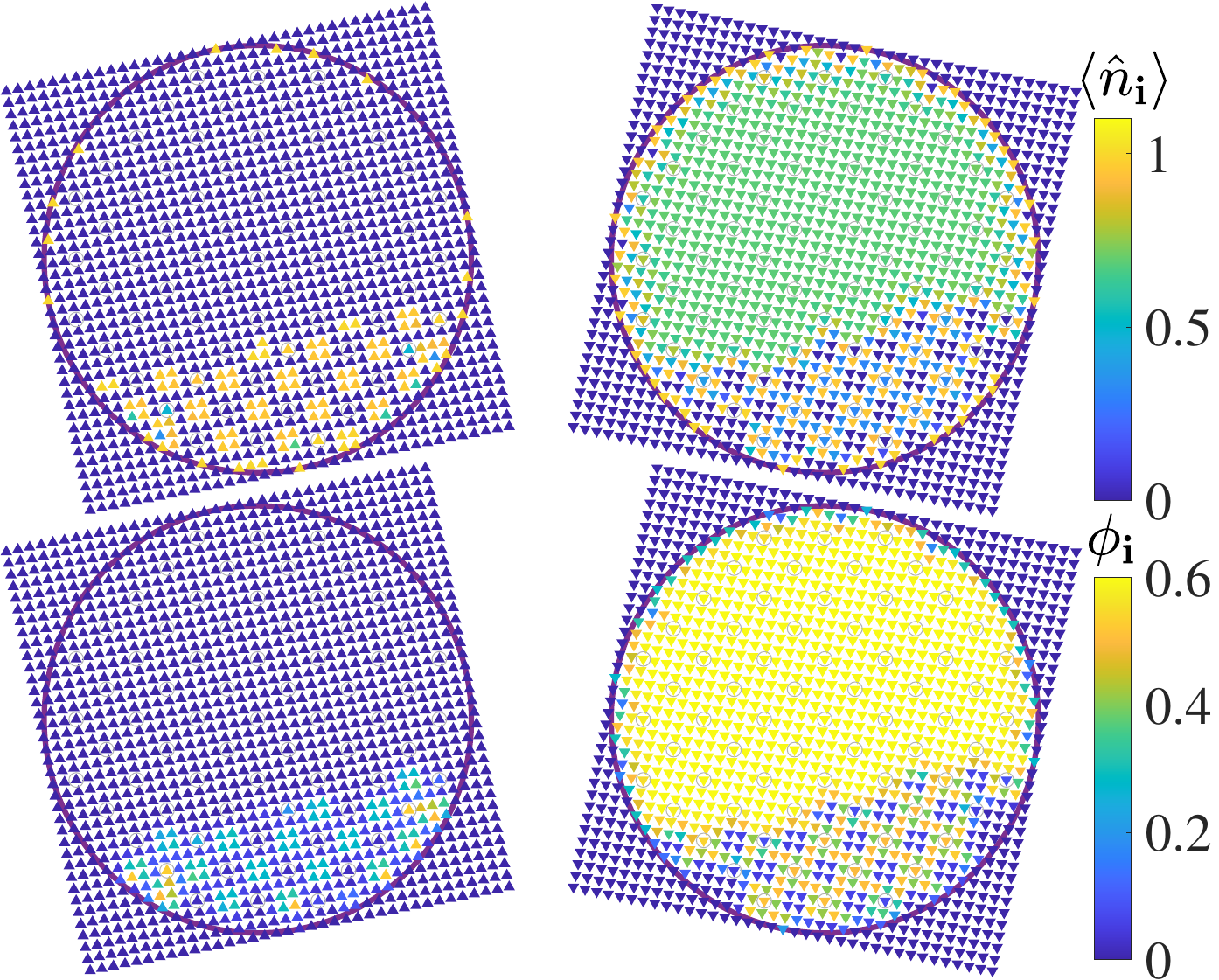} & \includegraphics[scale=0.15]{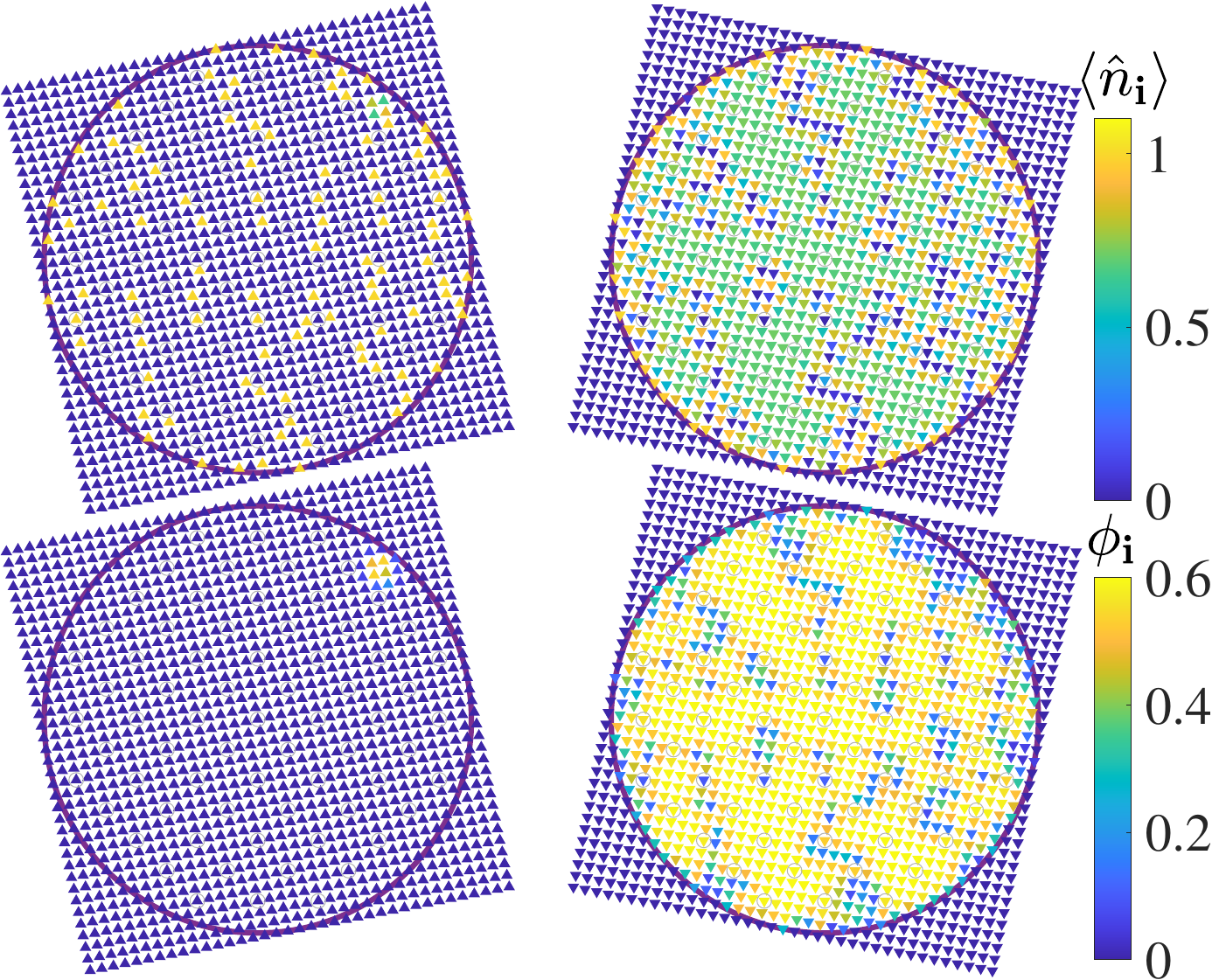} & \includegraphics[scale=0.15]{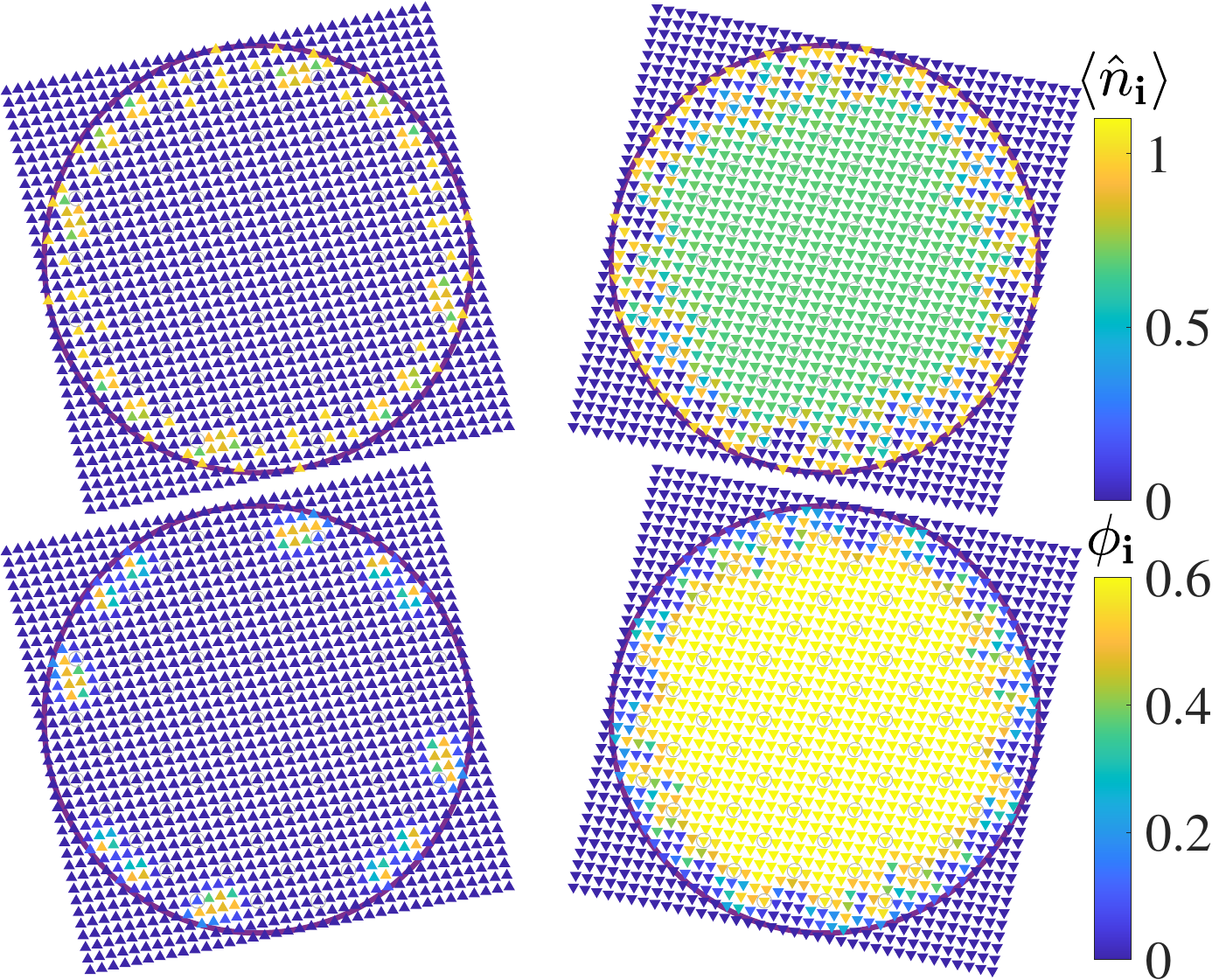}\tabularnewline
\hline 
$E=60.26U$ & $E=58.20U$ & $E=57.47U$ & $E=56.72U$\tabularnewline
\hline 
\end{tabular}
\par\end{centering}
\caption{Degenerate ground state and metastable state configurations for the
system parameter that corresponds to Fig.~3(d2) in the manuscript.}\label{tab:degenerate}
\end{table*}

\section{Finite-size effects}\label{sec:Finite_size}

{In this section, we present a detailed analysis of the finite-size
effects in our numerical simulations. Given the spatial dependence
of the interaction $U_{\mathbf{i}_{\uparrow}\mathbf{j}_{\downarrow}}(g_{\uparrow\downarrow},\theta)$
and the potential impact of the system size on the observed physical
properties, we have carried out additional simulations for larger
lattice sizes to ensure the reliability of our results.}

{To investigate the finite-size effects, we compare the phase diagrams
and typical physical quantities between two system sizes: $37\times37$
and $49\times49$. Specifically, we focus on the smallest twist angle
$\theta=7.63^{\circ}$ studied in our manuscript. Figs~\ref{fig:SystemSize}
(a) and (c) shows a direct comparison of the phase diagrams for the
$37\times37$ and $49\times49$ systems, and both phase diagrams display
qualitatively consistent results. In addition to the phase diagrams,
we also compare the spatial configurations of relevant physical observables,
such as the superfluid order parameter $\phi_{\mathbf{i}_{\sigma}}\equiv\langle\hat{b}_{\mathbf{i}_{\sigma}}\rangle$
and density distribution $\langle\hat{n}_{\mathbf{i}_{\sigma}}\rangle$.
Figs~\ref{fig:SystemSize} (b) and (d) display representative configurations
for both system sizes. The spatial profiles of the superfluid order
parameter and density exhibit negligible differences between the two
system sizes, further confirming that the 37\texttimes 37 system adequately
captures the essential physics of the system.}

\section{Spontaneously breaking of $C_{4}$ symmetry and metastable states}\label{sec:C4}

{In this section, we investigate the spontaneous breaking of $C_{4}$
rotational symmetry and the metastable states observed in our simulations.
These phenomena arise due to the intrinsic disorder potential, leading
to the four degenerate ground states and a rich set of metastable
states. To demonstrate the degeneracy of these ground states, we performed
simulations with different initial conditions for the system parameters
corresponding to Fig.~3(d2) of the main text. As shown in the first
row of the plots in Table~\ref{tab:degenerate}, we identified four
distinct ground state configurations possessing identical energy.
This confirms the spontaneous breaking of the $C_{4}$ symmetry in
the system.}

{In addition to the four degenerate ground states, our simulations
reveal the presence of numerous metastable configurations within the
Bose glass phase. These metastable states have higher energies than
the ground states. The second row of the plots in Table~\ref{tab:degenerate}
provides a representative set of metastable configurations for the
same parameter, illustrating the diversity of particle distributions
and their associated energies. This observation aligns with the expected
behavior of a Bose glass, which is typically characterized by the
coexistence of multiple metastable configurations.}

\begin{figure*}
\begin{centering}
\includegraphics[totalheight=2.1in]{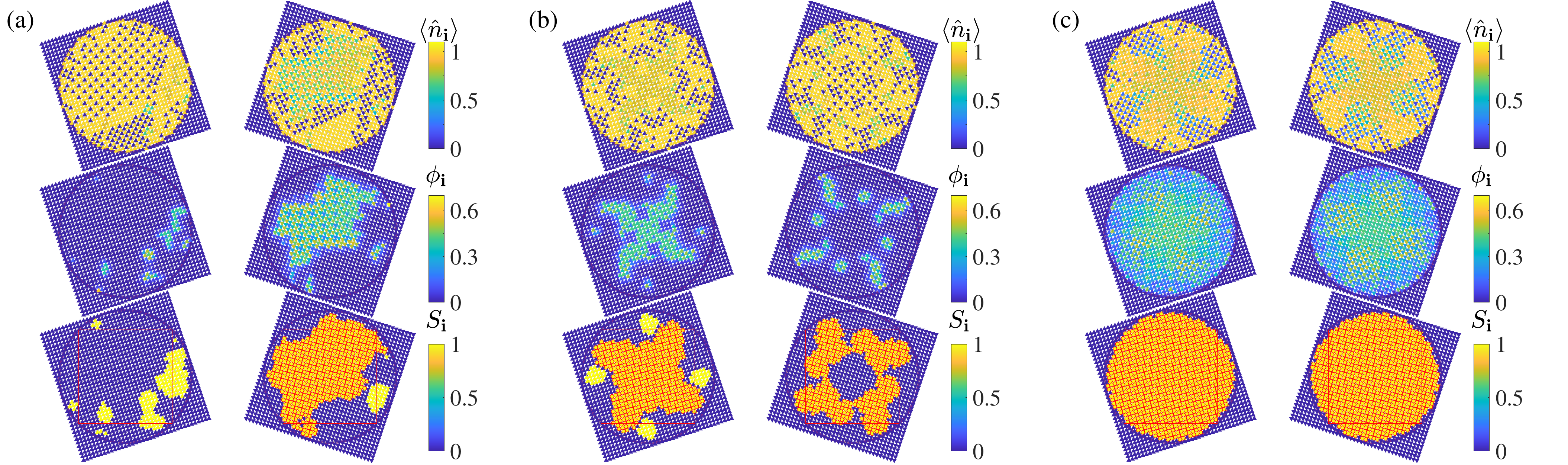}
\par\end{centering}
\caption{Identification of the Bose glass phase. (a) Typical configurations
of the Bose glass phase {[}induced Mott insulator phase with non-percolated
superfluid islands as well as the Fig.~2(c4) in the main text{]}.
(b) Typical configurations of the induced Mott insulator phase with
percolated superfluid region {[}Fig.~2(c3) in the main text{]}. (c)
Typical configurations of the superfluid phase {[}Fig.~2(c2) in the
main text{]}. The percolated superfluid regions across the length
scale of the system size (the red square) are marked by the small
red circles.}
\label{fig6}
\end{figure*}

\section{Bose glass identification}\label{sec:BG}

{To identify the Bose glass phase in our numerical simulations, we
implemented a systematic approach based on the percolation properties
of the superfluid regions and the spatial characteristics of the density
distribution. This section outlines the methodology used to identify
the Bose glass phase.}

{We use the approach outlined by references \cite{Niederle_2013BGIdentification_square,Johnstone_2021BGIdentification}
to analyze percolation of SF clusters on a discrete function $S_{\mathbf{i}_{\sigma}}$.
We label the lattice site $\mathbf{i}_{\sigma}$ with integer density
distribution $\langle\hat{n}_{\mathbf{i}_{\sigma}}\rangle$ and zero
SF order parameter $\phi_{\mathbf{i}_{\sigma}}$ as MI site with $S_{\mathbf{i}_{\sigma}}=0$,
and label the site with non-integer filling and nonzero $\phi_{\mathbf{i}_{\sigma}}$
as SF site with $S_{\mathbf{i}_{\sigma}}=1$. The BG phase is identified
by examining whether a percolated superfluid region exists on the
length scale of the system size. If none of the connected clusters
formed by the SF sites ($S_{\mathbf{i}_{\sigma}}=1$) percolates across
the system, the phase is identified as the BG phase. This phase is
characterized by isolated SF clusters embedded within an MI sea ($S_{\mathbf{i}_{\sigma}}=0$),
resulting in a non-vanishing density of SF sites that lack global
connectivity \cite{Niederle_2013BGIdentification_square}. Indeed,
in the typical BG phase observed in our numerical simulations, isolated
superfluid islands (i.e., non-percolated superfluid regions) are usually
surrounded by MI regions. The ``induced MI'' phase defined in the
main text is characterized by regions of the system where sites are
filled with an integer number of atoms and have no associated SF order
parameter. Consequently, the BG phase can still be considered a form
of ``induced MI'' as long as there are MI regions present. }

{Following the percolation identification approach on a square lattice
described in references \cite{Niederle_2013BGIdentification_square,Barman2013BGsquare},
we apply a similar method in our system with an external cylinder
box trap. Specifically, we inscribe the largest possible square within
the circular region to assess percolation. To ensure robustness, this
inscribed square is systematically rotated from $0{^\circ}$ to $360{^\circ}$,
allowing us to evaluate percolation for all orientations. Examples
of the identification of BG phase are presented Fig. \ref{fig6}.}

\end{document}